\documentclass[12pt,twoside]{article}
\usepackage{correl14}

\def\TheTitle{Making Use of Self-Energy Functionals: \newline The Variational Cluster Approximation}
\def\TheHeading{Variational Cluster Approximation}
\def\TheAuthor{Michael Potthoff}
\def\TheInstitute{I. Institut f\"ur Theoretische Physik \newline Universit\"at Hamburg}
\def\TheAddress{Jungiusstra{\ss}e 9, 20355 Hamburg, Germany}
\def\TheChapter{1}                       

\newcommand{\be}{\begin{equation}}
\newcommand{\ee}{\end{equation}}
\newcommand{\ba}{\begin{eqnarray}}
\newcommand{\ea}{\end{eqnarray}}
\newcommand{\ff}[1]{{\bm #1}}
\newcommand{\tr}{\mbox{tr}}
\newcommand{\Tr}{\mbox{Tr}}
\newcommand{\refeq}[1]{Eq.\ (\ref{eq:#1})}
\newcommand{\labeq}[1]{\label{eq:#1}}

\newcommand{\bi}{\begin{itemize}}
\newcommand{\ei}{\end{itemize}}

\newcommand{\ca}[1]{{\cal #1}}
\newcommand{\ew}[1]{\langle #1 \rangle}
\newcommand{\ket}[1]{| #1 \rangle}
\newcommand{\bra}[1]{\langle #1 |}

\begin{document}
\MakeTitle           

\section{Motivation}

\index{self-energy functional theory}

Self-energy-functional theory (SFT) \cite{Pot03a,PAD03,Pot11,Pot12} is a general theoretical frame that can be used to construct various approximate approaches by which the thermal properties and the spectrum of one-particle excitations of a certain class of correlated electron systems can be studied. 
The prototype considered here is the single-band Hubbard model \cite{Hub63,Gut63,Kan63} but, quite generally, the SFT applies to models of strongly correlated fermions on three or lower-dimensional lattices with local interactions.

There are several extensions of the theory, e.g.\ to systems with non-local interactions \cite{Ton05},
to bosonic systems \cite{KD05,KAvdL10a} and the Jaynes-Cummings lattice \cite{AHTL08,KAvdL10b}, 
to electron-phonon systems \cite{KMOH04},
to systems with quenched disorder \cite{PB07} 
as well as for the study of the real-time dynamics of systems far from thermal equilibrium \cite{HEAP13}. 
To be concise, those extensions will not be covered here.

\begin{figure}[b]
\includegraphics[width=0.85\textwidth]{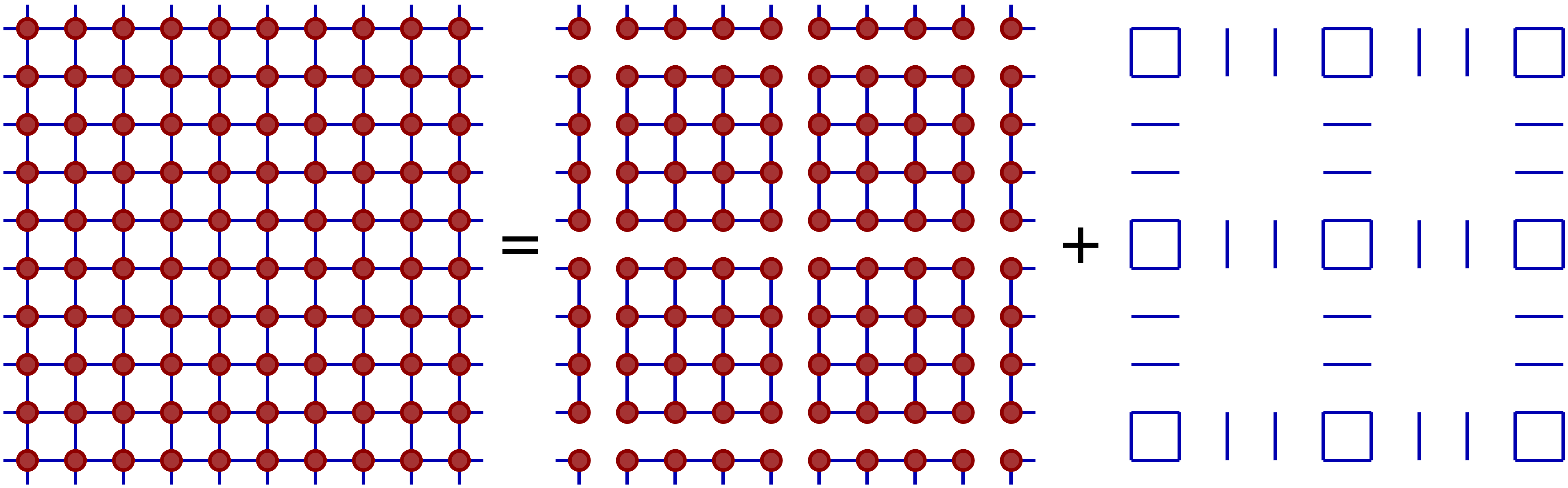}
\centering
\caption{
Sketch of the decomposition of the original system $H=H_{0}(\ff t) + H_{1}$ into a reference system $H'=H_{0}(\ff t') + H_{1}$ and the inter-cluster hopping $H_{0}(\ff V)$ for a square lattice and cluster size $L_{\rm c}=16$. Blue lines: nearest-neighbor hopping $t$. Red dots: on-site Hubbard interaction.
}
\label{fig:tiling}
\end{figure}

\index{variational cluster approximation}

The prime example of an approximation that can be constructed within the SFT is the variational cluster approximation (VCA) \cite{PAD03,DAH+04}. 
Roughly, one of the main ideas of the VCA is to adopt a ``divide and conquer strategy'':
A tiling of the original lattice into disconnected small clusters is considered, as shown in Fig.\ \ref{fig:tiling} for example. 
While the Hubbard model on the infinite square lattice cannot be solved exactly, there are no serious practical problems to the solve the same model for an isolated cluster or for a set of disconnected clusters.
The VCA constructs an approximate solution for the infinite lattice from the solution of the individual clusters by means of all-order perturbation theory for those terms in the Hamiltonian that connect the clusters. 

\index{cluster perturbation theory}

This is actually the concept of the so-called cluster perturbation theory (CPT) \cite{GV93,SPPL00}.
However, it is not sufficient in most cases, and we would like to go beyond the CPT.
The essential problem becomes apparent e.g.\ for a system with spontaneously broken symmetry such as an antiferromagnet. 
The antiferromagnetic state is characterized by a finite value for the sublattice magnetization which serves as an order parameter. 
On the other hand, quite generally, the order parameter must be zero for a system of finite size and thus for a small cluster in particular. 
Coupling finite (and thus necessarily non-magnetic) clusters by means of the CPT, however, one never gets to an antiferromagnetic solution for the infinite lattice. 
``Divide and conquer'' is not sufficient to describe the emergence of new phases with broken symmetries.

\index{spontaneous symmetry breaking}
\index{antiferromagnetism}

An obvious way out is to {\em enforce} a finite antiferromagnetic order parameter within each of the isolated clusters by applying a (staggered) magnetic field $B'$. 
Coupling those antiferromagnetic clusters may then result in an antiferromagnetic solution for the entire lattice. 

However, what determines the strength of this magnetic field?
As we are aiming at a description of {\em spontaneous} antiferromagnetic order, there is no external {\em physical} field $B$ that is applied to the original system ($B=0$). 
The field $B'$ is actually a ``Weiss field'', i.e.\ a fictitious field or ``mean field'' that is produced by the system itself. 
We are seeking for a formalism which allows for the formation of a finite Weiss field if this is ``favorable'', i.e.\ if a thermodynamical potential can be lowered in this way. 

\begin{figure}[t]
\includegraphics[width=0.35\textwidth]{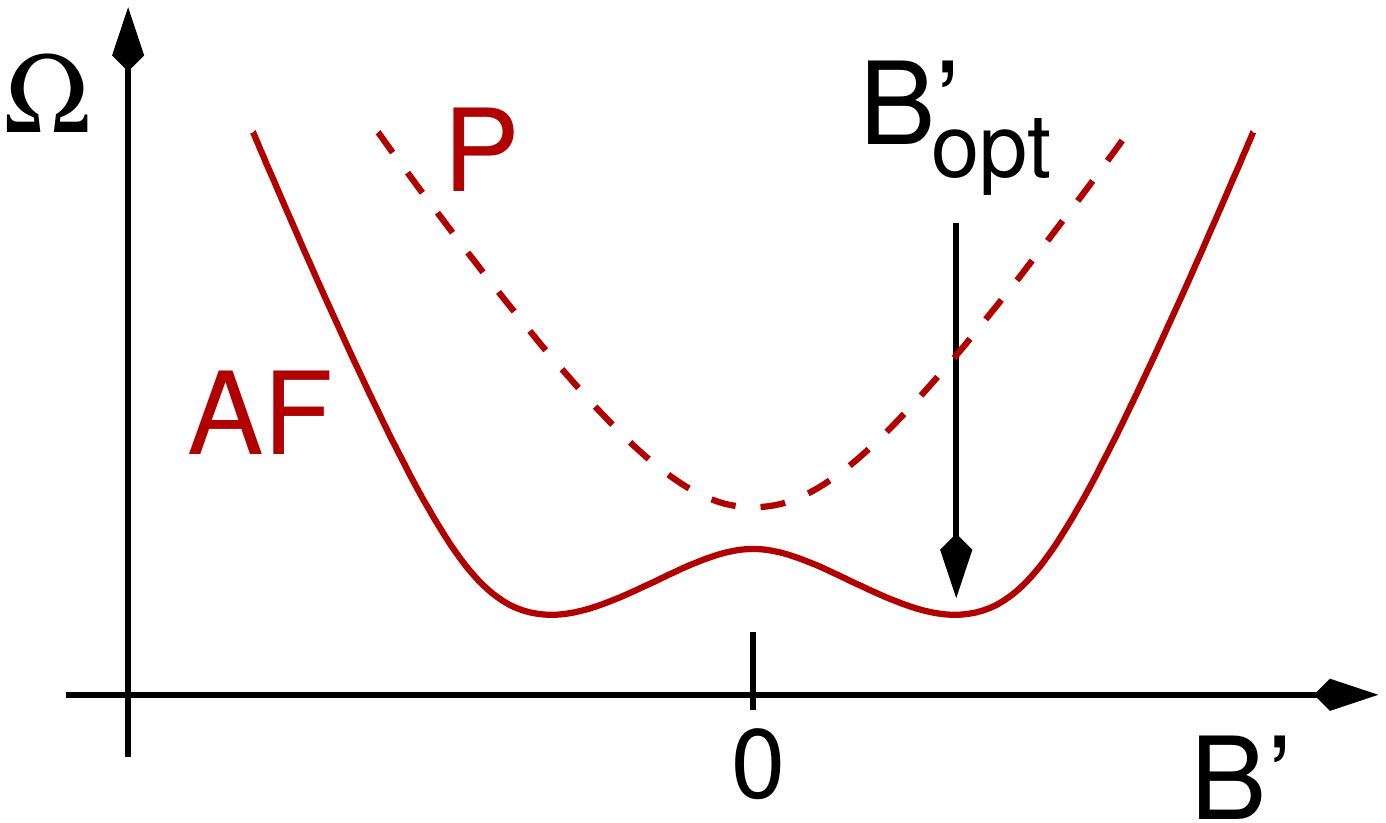}
\centering
\caption{
Grand potential $\Omega$ as a function of a Weiss field $B'$ in the case of a paramagnet (P) and in the case of an antiferromagnet (AF). $B'$ is a fictitious staggered field, the optimal value of which ($B'_{\rm opt}$) must be determined by minimization of $\Omega$. As there is no physically applied staggered field, i.e.\ $B=0$, a finite $B'_{\rm opt}$ indicates {\em spontaneous} symmetry breaking.
}
\label{fig:field}
\end{figure}

Self-energy-functional theory just provides a relation $\Omega(B')$ between the grand potential of the system $\Omega$ and the Weiss field $B'$ that can be used to fix the optimal value $B'_{\rm opt}$ of the staggered magnetic field by minimization (see Fig.\ \ref{fig:field}): 
\be
  \frac{\partial \Omega(B')}{\partial B'} \Bigg|_{B'=B'_{\rm opt}}\stackrel{!}{=} 0
\labeq{weiss}  
\ee

The purpose of this lecture is to show how this can be achieved in practice.
To this end we have to answer the following ``how to'' questions:
\begin{itemize}
\item
How can we solve the problem for an isolated cluster?
\item
With this at hand, how can we construct a solution for the problem on the infinite lattice?
\item
How can we construct the relation $\Omega(B')$ such that \refeq{weiss} determines $B'_{\rm opt}$?
\end{itemize}
Actually, there is no reason to consider only a staggered magnetic field as a Weiss field. 
Another goal is therefore to generalize the idea to arbitrary Weiss fields or to an arbitrary set of variational parameters $\ff \lambda'$ that characterize the isolated cluster and that are optimized via $\partial \Omega(\ff \lambda'_{\rm opt}) / \partial \ff \lambda' \stackrel{!}{=} 0$.
Finally, the VCA should be compared with other theories available and the practical as well as principal limitations have to be discussed.

\section{The cluster approach}

\subsection*{Tiling the lattice into small clusters}

We start with the second question and consider a simple non-interacting system given by
\be
  H_{0} = \sum_{ij\sigma} t_{ij} c_{i\sigma}^{\dagger} c_{j\sigma} = H_{0}(\ff t) \: .
\labeq{h0}
\ee
Here, $c_{i\sigma}^{\dagger}$ creates an electron with spin $\sigma=\uparrow, \downarrow$ at the site $i$ of a $D$-dimensional lattice, and $t_{ij}$ are the (spin-independent) hopping parameters which are also considered as the elements of the hopping matrix $\ff t$. 
Furthermore, 
\be
  H'_{0} = \sum_{ij\sigma} t'_{ij} c_{i\sigma}^{\dagger} c_{j\sigma} = H_{0}(\ff t') \: ,
\labeq{h0p}  
\ee
denotes the Hamiltonian of the system with decoupled clusters (see Fig.\ \ref{fig:tiling} and take $H_{1}=0$). 
If $L$ is the number of lattice sites in the original lattice model $H_{0}$ and $L_{\rm c}$ is the number of sites in an individual cluster, there are $L/L_{\rm c}$  decoupled clusters.
We assume that all clusters are identical.
In terms of hopping matrices, we have 
\be
\ff t = \ff t' + \ff V
\labeq{decomp}
\ee
where $\ff V$ is the inter-cluster hopping. 

\index{Green's function}

Consider the resolvent of the hopping matrix, i.e.\ the Green's function
\be
  \ff G_{0}(\omega) = \frac{1}{\omega + \mu - \ff t} \: .
\label{eq:g0}
\ee
Here, $\omega$ is a complex frequency (units with $\hbar = 1$ are used). 
We have also introduced the chemical potential $\mu$ (which is not important here but used later).
Furthermore, we employ a matrix notation and write $\omega$ rather than $\omega \ff 1$ for short etc.
Note the $(\cdots)^{-1}$ and $1/(\cdots)$ means matrix inversion. 

Having the Green's function of the reference system at hand, 
\be
  \ff G_{0}'(\omega) = \frac{1}{\omega + \mu - \ff t'} \: ,
\ee
how can be get the Green's function of the original model? With some algebra, one easily derives the equation
\be
  \ff G_{0}(\omega) = \ff G_{0}'(\omega) + \ff G_{0}'(\omega) \ff V \ff G_{0}(\omega)
\label{eq:cpt0}
\ee
which is solved by 
\be
  \ff G_{0}(\omega) = \frac{1}{\ff G_{0}'(\omega)^{-1} - \ff V} \: .
\ee
We see that using Green's functions it is formally rather easy to couple a system of isolated clusters.

\subsection*{Cluster perturbation theory}

\index{cluster perturbation theory}

Actually, we are interested in interacting systems. 
For the single-orbital model $H_{0}$, the only possible local interaction is a Hubbard interaction of the form
\be
  H_{1} = \frac{U}{2} \sum_{i\sigma} n_{i\sigma}  n_{i-\sigma} \; 
\labeq{h1}
\ee
with $n_{i\sigma} = c_{i\sigma}^{\dagger} c_{i\sigma}$ and where $U$ is the interaction strength.
Since $H_{1}$ is completely local, the Hamiltonian of the so-called ``reference system'' $H' = H_{0}(\ff t') + H_{1}$ is obtained form the Hamiltonian of the ``original system'' $H = H_{0}(\ff t) + H_{1}$ by switching off the inter-cluster hopping $\ff V$.

For a small cluster and likewise for a system of disconnected clusters, even for the interacting case, it is comparatively simple to solve the problem exactly (by numerical means if necessary) while for the original lattice model this is a hard problem.
One therefore cannot expect a simple relation between the original and the reference system like Eq.\ (\ref{eq:cpt0}).
Nevertheless, as it is too tempting, we will write down
\be
  \ff G(\omega) = \ff G'(\omega) + \ff G'(\omega) \ff V \ff G(\omega)
\label{eq:cpt}
\ee
where now $\ff G$ and $\ff G'$ are interacting Green's functions.
This is an equation that constitutes the cluster-perturbation theory \cite{GV93,SPPL00}.
It must be seen as an approximate way to compute the Green's function of the interacting model from the exact cluster Green's function.
In a way the approximation is controlled by the size $L_{\rm c}$ of the clusters in the reference system since for $L_{\rm c} \to \infty$ one can expect the approximation to become exact.
In fact, the CPT is not too bad and has been successfully applied in a couple of problems, see Ref.\ \cite{Sen12} and references therein.

\subsection*{Green's function and exact diagonalization}
\label{sec:green}

\index{Green's function}
\index{exact diagonalization}

Before proceeding with the interpretation of the CPT equation (\ref{eq:cpt}) which provides an approximate expression for $\ff G(\omega)$, let us give the {\em exact} definition of the Green's function for the interacting case.
Its elements are defined as 
\be
  G_{ij\sigma}(\omega) = \int_{-\infty}^{\infty} dz \, \frac{A_{ij\sigma}(z)}{\omega - z} \: , 
\label{eq:green}
\ee
where $\omega$ is an arbitrary complex frequency and where 
\be
  A_{ij\sigma} (z) = \int_{-\infty}^{\infty} dt \, e^{{izt}} A_{ij\sigma}(t)
\label{eq:afourier}
\ee
is the single-particle spectral density whose Fourier transform 
\be
  A_{ij\sigma}(t) = \frac{1}{2\pi} \langle [ c_{i\sigma}(t) , c_{j\sigma}^{\dagger}(0) ]_{+} \rangle
\label{eq:aoft}
\ee
is given as the thermal expectation value of the anti-commutator of the annihilator with the creator in the (grand-canonical) Heisenberg picture, e.g.
\be
  c_{i\sigma}(t) = e^{i (H - \mu N) t} c_{i\sigma} e^{-i (H - \mu N) t}
\ee
with $N=\sum_{i\sigma} n_{i\sigma} = \sum_{i\sigma} c_{i\sigma}^{\dagger} c_{i\sigma}$.
The thermal average is a grand-canonical average
$\langle \cdots \rangle = Z^{-1} \tr (e^{-\beta (H-\mu N)}\cdots )$
where $Z=Z(\beta,\mu) =  \tr \, e^{-\beta (H-\mu N)}$ is partition function at the chemical potential $\mu$ and the inverse temperature $\beta$.

In the case of a non-interacting system, one may use the Baker-Campbell-Hausdorff formula to get the simple result
($\ff t$: hopping, $t$: time):
\be
  c_{i\sigma}(t) = \sum_{j} \left( e^{- i (\ff t - \mu) t} \right)_{ij} c_{j\sigma} 
\ee
which can be used in Eq.\ (\ref{eq:aoft}), and then via the Fourier transformation (\ref{eq:afourier}) and finally the Hilbert transformation (\ref{eq:green}) one arrives at the result given by Eq.\ (\ref{eq:g0}) above. 

In the interacting case ($U>0$), one may compute the Green's function from the eigenvalues $E_{n}$ and eigenstates $|n\rangle$ of the (grand-canonical) Hamiltonian: 
\be
  (H - \mu N) | n \rangle = E_{n} | n \rangle \: .
\label{eq:eigen}  
\ee
Using a resolution of the unity, $\ff 1 = \sum_{n} | n \rangle \langle n |$, in Eq.\ (\ref{eq:aoft}), one can easily do the calculation and arrives at 
\be
  G_{ij\sigma}(\omega) = \frac{1}{Z} \sum_{mn} \frac{(e^{-\beta E_{m}} + e^{-\beta E_{n}})
  \langle m | c_{i\sigma} | n \rangle \langle n | c_{j\sigma}^{\dagger} | m \rangle
  }{\omega - (E_{n} - E_{m})} \: .
\labeq{lehmann}
\ee
However, as one must solve the many-body energy eigenvalue problem (\ref{eq:eigen}), this way to calculate the Green's function is obviously impossible in practice for the Hamiltonian of the original system -- the Hilbert-space dimension exponentially increases with $L$.
On the other hand, for the reference system and if the size of the cluster $L_{\rm c}$ is not too large, this can be done numerically.
For a half-filled system ($N=L_{\rm c}$), up to $L_{\rm c}=8$ sites can be managed in this way easily. 
At zero temperature, using the Lanczos algorithm \cite{Koc11}, the Green's function for somewhat larger clusters can be computed, typically $L_{\rm c}\le 12$ at half-filling.
This already answers the first question posed in the introduction.

\subsection*{Freedom in the CPT construction}

The CPT gives a preliminary answer to the second question. 
However, it is easily seen that the answer is not unique: 
Consider a ``modified'' reference system with a Hamiltonian
\be
  H_{0}(\widetilde{\ff t'}) + H_{1} = H_{0}(\ff t') + H_{0}(\Delta \ff t) + H_{1} = \sum_{ij\sigma} (t'_{ij} + \Delta t_{ij}) c_{i\sigma}^{\dagger} c_{j\sigma} + H_{1} \; ,
\ee
i.e.\ a reference system where $\ff t' \mapsto \widetilde{\ff t'} = \ff t' + \Delta \ff t'$.
The new reference system shall still describe the same set of decoupled clusters but with different intra-cluster hoppings $\widetilde{\ff t'}$.
The modified non-interacting Green's function of the reference system is $\widetilde{\ff G}_{0}'(\omega) = 1 / (\omega + \mu - \widetilde{\ff t'})$. 
Now, the non-interacting Green's function of the original model is obtained from the equation
\be
  \ff G_{0}(\omega) = \widetilde{\ff G}'(\omega) + \widetilde{\ff G}'(\omega) \widetilde{\ff V} \ff G_{0}(\omega)
\label{eq:cpt1}
\ee
with the modified inter-cluster hopping $\widetilde{\ff V} = \ff t - \widetilde{\ff t'} = \ff V - \Delta \ff t'$.
$\ff G_{0}(\omega)$ can be considered as the limit of a geometrical series that is found by iterating equation (\ref{eq:cpt1}):
\be
  \ff G_{0}(\omega) = \widetilde{\ff G}'(\omega) + \widetilde{\ff G}'(\omega) \widetilde{\ff V} \widetilde{\ff G}'(\omega) + \cdots \; . 
\ee
We infer that $\ff G_{0}(\omega)$ can be obtained by (all-order) perturbation theory in $\widetilde{\ff V}$ when expanding around the Green's function of the modified reference system given by the hopping matrix $\widetilde{\ff t'}$. 
Obviously, the same result is obtained by perturbation theory in $\ff V$ around the Green's function of the modified reference system with hopping matrix ${\ff t'}$. 
This freedom in choosing the starting point for perturbation theory that we have in the non-interacting case turns into a real problem for the interacting case. 
Namely, since the CPT equation (\ref{eq:cpt}) is approximate, we generally have:
\be
  \widetilde{\ff G}(\omega) 
  \equiv 
  \widetilde{\ff G}'(\omega) + \widetilde{\ff G}'(\omega) \widetilde{\ff V} \widetilde{\ff G}'(\omega) + \cdots 
  \ne
  {\ff G}'(\omega) + {\ff G}'(\omega) {\ff V} \ff G'(\omega) + \cdots 
  \equiv 
  {\ff G}(\omega) \: .
\ee
Concluding, different starting points, $\ff t'$ and $\widetilde{\ff t'}$, for the all-order cluster perturbation theory, in $\ff V$ and $\widetilde{\ff V}$, lead to different results, ${\ff G}(\omega)$ and $\widetilde{\ff G}(\omega)$, respectively.

But which is the ``right'' starting point?
The idea is to turn the problem into an advantage by ``optimizing'' the starting point: 
This can be done by making use of a variational principle, i.e.\ by expressing a thermodynamical potential, e.g.\ the grand potential $\Omega$, as a function of $\ff t'$ and by subsequent minimization. 
The optimal $\ff t'_{\rm opt}$ shall be obtained by 
\be
  \frac{\partial \Omega(\ff t')}{\partial \ff t'} \Bigg|_{\ff t'=\ff t'_{\rm opt}}\stackrel{!}{=} 0 \: .
\ee
We see that the set of variational parameters is just the set of hopping parameters of the reference systems or, in the case of multi-orbital models, simply the set of {\em all} one-particle parameters except for those, of course, that would couple the different clusters. 
This set also includes a staggered magnetic field
\be
  H_{0}(\widetilde{\ff t'}) 
  =
  H_{0}({\ff t'}) 
  -
  B' \sum_{i} z_{i} (n_{i\uparrow} - n_{i\downarrow}) \: ,
\ee
where $z_{i} = \pm 1$ alternates between the sites of a bipartite lattice.

\subsection*{The Ritz principle?}

\index{Ritz variational principle}

The most popular variational principle is the Ritz variational principle. 
It states that
\be
  E[\ket{\Psi}] = \langle \Psi  | H | \Psi \rangle = \mbox{min.}
\ee
for the ground state of $H$ when the search extends over all normalized trial states $\langle \Psi | \Psi \rangle = 1$.
Evaluated at the ground state $| \Psi_{0} \rangle$, the functional yields the ground-state energy $E[\ket{\Psi_{0}}] = E_{0}$.

Hence, a straightforward idea that suggests itself is to compute the normalized ground state $| \Psi(\ff t') \rangle$ of a reference system with hopping matrix $\ff t'$ and to use this as a trial state. 
The trial state can be varied by varying the parameters $\ff t'$, and the optimal parameters are given by 
\be
  \frac{\partial E[\ket{\Psi(\ff t')}]}{\partial \ff t'} \Bigg|_{\ff t'=\ff t'_{\rm opt}}\stackrel{!}{=} 0 \: .
\ee
To test this idea, let 
\be
\ket{\Psi(\ff t')} = \ket{\Psi_{1}(\ff t'_{1})} \otimes \ket{\Psi_{2}(\ff t'_{2})} \otimes \cdots \otimes \ket{\Psi_{L/L_{\rm c}}(\ff t'_{L/L_{\rm c}})}
\ee
be the ground state of $H' = H_{0}(\ff t') + H_{1}$.
It is given as a product of the ground states of the $L/L_{\rm c}$ individual clusters where the ground state of the $I$-th cluster with hopping matrix $\ff t'_{I}$ is 
$\ket{\Psi_{I}(\ff t'_{I})}$.
Now, if $E_{0}(\ff t')$ denotes the ground-state energy of the reference system, 
\be
  E[\ket{\Psi(\ff t')}] 
  =
  \langle \Psi(\ff t') | (H_{0}(\ff t') + H_{0}(\ff V) + H_{1}) | \Psi(\ff t') \rangle
  =
  E_{0}(\ff t') + \langle \Psi(\ff t') | H_{0}(\ff V) | \Psi(\ff t') \rangle \: .
\ee
However, the inter-cluster hopping Hamiltonian $H_{0}(\ff V)$ only contains terms like $c_{i\sigma}^{\dagger} c_{j\sigma}$ where the sites $i$ and $j$ belong to {\em different} clusters, say $I$ and $J$. 
Hence, 
$\langle \Psi(\ff t') | c_{i\sigma}^{\dagger} c_{j\sigma} | \Psi(\ff t') \rangle
=
\bra{\Psi_{I}(\ff t'_{I})} \otimes \bra{\Psi_{J}(\ff t'_{J})} 
c_{i\sigma}^{\dagger} c_{j\sigma} 
\ket{\Psi_{J}(\ff t'_{J})} \otimes \ket{\Psi_{I}(\ff t'_{I})} 
=
\bra{\Psi_{I}(\ff t'_{I})} 
c_{i\sigma}^{\dagger}
\ket{\Psi_{I}(\ff t'_{I})}
\bra{\Psi_{J}(\ff t'_{J})} 
c_{j\sigma} 
\ket{\Psi_{J}(\ff t'_{J})}  
=0
$
as enforced by the conservation of the total particle number. 
This means that we are left $E[\ket{\Psi(\ff t')}]  = E_{0}(\ff t')$. 
As this implies that the optimal parameters $\ff t'_{\rm opt}$ do not at all depend on $\ff V$, the result is trivial and useless, unfortunately.
Even worse, the Hellmann-Feynman theorem \cite{Fey39} tells us that
\be
  \frac{\partial}{\partial \ff t'}
  E[\ket{\Psi(\ff t')}]
  = 
  \frac{\partial}{\partial \ff t'}
  E_{0}(\ff t')
  = 
  \frac{\partial}{\partial \ff t'}
  \bra{\Psi(\ff t')}
  (H_{0}(\ff t') + H_{1})
  \ket{\Psi(\ff t')}
  =  
  \bra{\Psi(\ff t')}
  \frac{\partial H_{0}(\ff t')}{\partial \ff t'} 
  \ket{\Psi(\ff t')} \: .
\ee
This means that, using the Ritz principle, the variational parameters should be determined such that all one-particle intra-cluster correlation functions $\langle c^{\dagger} c \rangle$, in addition to the inter-cluster correlation functions, vanish. 

Concluding, optimizing cluster-perturbation theory cannot be done with the help of the Ritz principle. 
We mention in passing that this also holds for its finite-temperature and mixed state generalization 
\cite{Gib48,Fey55}
\be
\Omega[\rho] = \tr \Big( 
\rho ( H - \mu N + T \ln \rho )
\Big) \stackrel{!}{=} \mbox{min.} 
\: , 
\labeq{finite-t}
\ee
where the grand potential, expressed as a functional of the density matrix, is at a minimum for the thermal density matrix 
$\rho = \exp(-\beta (H - \mu N)) / \tr \exp(-\beta (H - \mu N))$. 
While this is an extremely useful variational principle, it cannot be used here:
A trial density matrix $\rho(\ff t')$, defined as the thermal density matrix of a reference system with a hopping matrix $\ff t'$ that describes decoupled clusters, is a simple product of individual cluster density matrices only. 
As for the standard Ritz principle, this implies that inter-cluster one-particle correlations are neglected altogether.

\section{Diagrammatic perturbation theory}

\index{many-body perturbation theory}

\subsection*{$S$-matrix and Green's function}

As we have already seen, Green's functions, opposed to wave functions or density matrices, can be used to couple isolated clusters. 
All-order perturbation theory in the inter-cluster hopping $\ff V$ yields the exact Green's function in the non-interacting ($U=0$) case and an approximate (CPT) Green's function for $U>0$. 
For the necessary optimization of the starting point, i.e.\ of the intra-cluster one-particle parameters $\ff t'$, we should therefore try to formulate a variational principle based on Green's functions, i.e.\ a principle of the form $\delta \Omega[\ff G(\omega)] / \delta \ff G(\omega) \stackrel{!}{=} 0$, and try ``test Green's functions'' $\ff G'(\omega)$ taken from the reference system.
In fact, a variational principle of this type can be constructed with the help of all-order perturbation theory in $U$ \cite{LW60,AGD64}.
Vice versa, a systematic and general perturbation theory in $U$ (and also in $\ff V$) requires to put Green's functions in the focus of the theory.
Here, a brief sketch is given only, details can be found in Refs.\ \cite{AGD64,FW71,NO88}, for example.
Our goal is to use diagrammatic perturbation theory as a ``language'' that can be used to formulate a Green's-function-based variational principle. 

\index{diagram technique}

We decompose the (grand-canonical) Hamiltonian $\ca H \equiv H - \mu N$ into a ``free'' part $\ca H_{0} = H_{0} - \mu N$ and the interaction $H_{1} \equiv \ca H - \ca H_{0}$.
Next we define, for $0\le \tau, \tau' \le \beta$, the so-called S-matrix as
\be
  S(\tau, \tau') = e^{\ca H_0 \tau} e^{- \ca H (\tau - \tau')} e^{-\ca H_0\tau'} \; ,
\labeq{sm}
\ee
One may interpret $\tau = i t$ as an ``imaginary-time'' variable (where $t$ is real). 
This ``Wick rotation'' in the complex time plane has the formal advantage that the thermal density matrix, $\propto e^{-\beta \ca H}$, is just given by the time-evolution operator, $e^{-i\ca Ht} = e^{-\ca H \tau}$ at $\tau = \beta$.

There are two main purposes of the S-matrix.
First, it can be used to rewrite the partition function in the following way:
\be
Z
= 
\tr \, e^{-\beta \ca H}
=
\tr \left(
 e^{-\beta \ca H_0 }
 e^{\beta \ca H_0}
 e^{-\beta \ca H}
\right)
= 
\tr \left(
 e^{-\beta \ca H_0}
 S(\beta,0)
\right)
= 
 Z_{0}
 \ew{
 S(\beta,0)
 }^{(0)} \: .
\ee
The partition function of the interacting system is thereby given in terms of the partition function of the {\em free} system, which is known, and a {\em free} thermal expectation value of the S-matrix.
The second main purpose is related to the imaginary-time Green's function which, for $-\beta < \tau < \beta$, is defined via
\be
G_{ij\sigma}(\tau) 
=
-
\langle
\ca T
c_{i\sigma}(\tau) c_{j\sigma}^\dagger(0)
\rangle
\ee 
in terms of an annihilator and a creator with imaginary Heisenberg time dependence:
\be
  c_{i\sigma}(\tau) = e^{\ca H \tau} c_{i\sigma} e^{-\ca H \tau} \: , \qquad 
  c^{\dagger}_{j\sigma}(\tau) = e^{\ca H \tau} c^{\dagger}_{j\sigma} e^{-\ca H \tau} \: .
\ee
Furthermore, $\ca T$ is the (imaginary) time-ordering operator.
With the help of the S-matrix the interacting time dependence can be transformed into a {\em free} time dependence, namely:
\be
c_{i\sigma}(\tau)  
= 
S(0,\tau) c_{I,i\sigma}(\tau) S(\tau,0)
\; , \qquad
c^\dagger_{j\sigma}(\tau)  
= 
S(0,\tau) c^\dagger_{I,j\sigma}(\tau) S(\tau,0) \: .
\ee 
Here, the index $I$ (``interaction picture'') indicates that the time dependence is due to $\ca H_0$ only. 
This time dependence is simple and can be derived with the Baker-Campbell-Hausdorff formula again:
\be
  c_{I,i\sigma}(\tau)
  =
  \sum_{j}
  \left(
  e^{-(\ff t -\mu)\tau}
  \right)_{ij}
  c_{j\sigma}
  \; , \qquad
  c^\dagger_{I,i\sigma}(\tau)
  =
  \sum_{j}
  \left( 
  e^{+(\ff t -\mu)\tau}
  \right)_{ij}
  c^\dagger_{j\sigma} \: .
\ee
Outside the imaginary-time interval $-\beta < \tau < \beta$, the Green's function is defined as the periodic continuation: $G_{ij\sigma}(\tau + k \cdot 2 \beta) = G_{ij\sigma}(\tau)$ for any integer $k$.
This function has a discrete Fourier representation: 
\be
  G_{ij\sigma}(\tau) = \frac{1}{\beta} \sum_{n=-\infty}^\infty G_{ij\sigma}(i\omega_n)  
  \, e^{-i \omega_n \tau} \; ,
\ee
where the Fourier coefficients $G_{ij\sigma}(i\omega_n)$ are defined at the so-called fermionic Matsubara frequencies $i\omega_{n} = i (2n+1) \pi / \beta$ for integer $n$ and can be computed from $G_{ij\sigma}(\tau)$ as 
\be
  G_{ij\sigma}(i\omega_n) = \int_0^\beta d\tau \, G_{ij\sigma}(\tau) \, e^{i\omega_n \tau} \: .
\ee
The Green's function $G_{ij\sigma}(\tau)$ is just a different representation of the Green's function $G_{ij\sigma}(\omega)$ introduced with Eq.\ (\ref{eq:green}) as its Fourier coefficients are given by $G_{ij\sigma}(i\omega_{n}) = G_{ij\sigma}(\omega)|_{\omega=i\omega_{n}}$.

The remaining problem consists in finding a much more suitable representation of the S-matrix. 
From its definition one straightforwardly derives the following equation of motion:
\be
  - \frac{\partial}{\partial \tau} S(\tau,\tau') = H_{1,I}(\tau) S(\tau,\tau') \: .
  \labeq{dgl}
\ee
Here, the time dependence of $H_{1,I}(\tau)$ is due to $H_0$ only. 
A formal solution of this differential equation with the initial condition $S(\tau, \tau) = 1$ can be derived easily using the time-ordering operator $\ca T$ again:
\be
  S(\tau,\tau') = \ca T \exp \left( - \int_{\tau'}^\tau \, d\tau'' H_{1,I}(\tau'') \right) \: .
\labeq{sr}
\ee
Note, that if all quantities were commuting, the solution of \refeq{dgl} would trivially be given by \refeq{sr} without $\ca T$. 
The appearance of $\ca T$ can therefore be understood as necessary to enforce commutativity.

Using this S-matrix representation, the partition function and the Green's function can be written as:
\be
\frac{Z}{Z_0}
=
 \Big\langle
 \ca T \exp \left( - \int_{0}^\beta \, d\tau'' H_{1,I}(\tau'') \right)
 \Big\rangle^{(0)}
\labeq{p1}
\ee
and
\be
G_{ij\sigma}(\tau)
=
- 
\frac{ \Big\langle 
\ca T \exp \left( - \int_{0}^\beta \, d\tau H_{1,I}(\tau) \right)
c_{I, i\sigma}(\tau) c_{I, j\sigma}^\dagger(0) \Big\rangle^{(0)} }
{
\Big\langle
 \ca T \exp \left( - \int_{0}^\beta \, d\tau H_{1,I}(\tau) \right)
 \Big\rangle^{(0)}
} \: .
\labeq{p2}
\ee
The important point is that the expectation values and time dependencies appearing here are free and thus known. 
Therefore, expanding the exponentials in \refeq{p1} and \refeq{p2} provides an expansion of the partition function and of the Green's function in powers of the interaction strength. 
The coefficients of this expansion are given as free expectation values of time-ordered products of annihilators and creators with free time dependencies.
In $k$-th order, this is a $k$-particle free correlation function which can be simplified by using Wick's theorem.
This is the central theorem of diagrammatic perturbation theory and applies to {\em free} higher-order correlation functions.

Consider the case of the partition function, for example.
At $k$-th order, the coefficient is given by a sum of $(2k)!$ terms, each of which factorizes into an $k$-fold product of terms of the form $\langle \ca T c_{i\sigma}(\tau) c^\dagger_{j\sigma}(\tau') \rangle^{(0)}$ called propagators.
Apart from a sign, propagator is nothing but the free Green's function. 
The summation of the $(2k)!$ terms is organized by means of a diagrammatic technique where ``vertices'' are linked via ``propagators''.
Wick's theorem and the details of technique can be found in Refs.\ \cite{AGD64,FW71,NO88}, for example.

\subsection*{Scattering at the inter-cluster potential, diagrammatically}
\label{sec:scatter}

\index{diagram technique}
\index{Feynman diagrams}

\begin{figure}[t]
\includegraphics[width=0.5\textwidth]{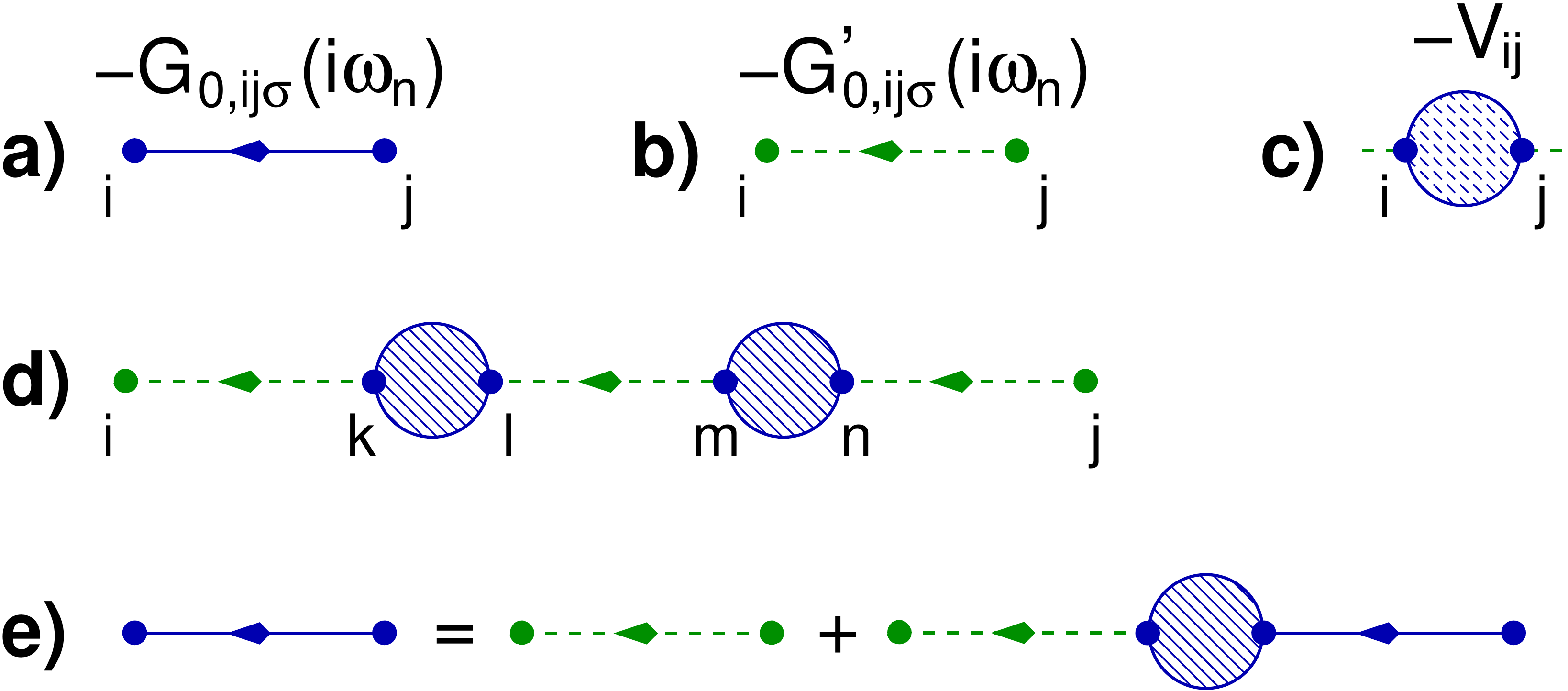}
\centering
\caption{
Diagrams for potential scattering. See text for discussion.
}
\label{fig:scatter}
\end{figure}

Here, it is sufficient to illustrate the technique. 
To this end, we first consider the simple and exactly solvable system that is given by the Hamiltonian $\ca H_{0}(\ff t) = H_{0}(\ff t) - \mu N$ (see \refeq{h0}). 
We decompose the Hamiltonian into a ``free'' part $\ca H_{0}(\ff t') = H_{0}(\ff t') - \mu N$  (see \refeq{h0p}) and an ``interaction'' $H_{1} \equiv H_{0}(\ff V)$ (see \refeq{decomp}).
The ``fully interacting'' propagator, which we are interested in, is $-G_{0,ij\sigma}(i\omega_{n})$ and is represented by an oriented line which starts at site $j$ where the electron is created ($c_{j\sigma}^{\dagger}$) and ends at site $i$ 
(see Fig.\ \ref{fig:scatter}a). 
The ``free'' propagator $-G'_{0,ij\sigma}(i\omega_{n})$ is represented by a dashed line
(see Fig.\ \ref{fig:scatter}b). 
Propagators carry a frequency $i\omega_{n}$ and a spin $\sigma$.
A circle with two links, one for an incoming and one for an outgoing propagator, is called a vertex and stands for the ``interaction'' $-V_{ij}$ itself
(see Fig.\ \ref{fig:scatter}c). 
According to Wick's theorem, the contribution of order $k$ to $-G_{0,ij\sigma}$ is obtained by drawing all topologically different diagrams where all links at $k$ vertices are connected by free propagators, except for two external links at the sites $i$ and $j$
(see Fig.\ \ref{fig:scatter}d). 
This contribution is calculated by performing the sums over ``internal'' variables (such as $k,l,m,n$ in Fig.\ \ref{fig:scatter}d) and respecting frequency and spin conservation at each vertex.
These diagram rules can be derived strictly by expanding \refeq{p2} and applying Wick's theorem. 
Together with \refeq{p1} this also leads to the important so-called linked-cluster theorem which allows us to concentrate on {\em connected} diagrams only.
The disconnected diagrams for the Green's function (i.e.\ with external links) exactly cancel diagrammatic contributions from the denominator in \refeq{p2}. 
As concerns closed diagrams (no external links) contributing to the partition function, \refeq{p1}, the sum of only the connected closed diagrams yields, apart from a constant, $\ln Z$, i.e.\ the grand potential.
For the simple case of scattering at the inter-cluster potential discussed at the moment, there is a single connected diagram at each order $k$ only, and thus the ``interacting'' Green's function is given by 
\be
  - G_{0,ij\sigma}(i\omega_{n}) 
  = 
  - G'_{0,ij\sigma}(i\omega_{n})
  +
  \sum_{kl}
  [- G'_{0,ik\sigma}(i\omega_{n})] \, [-V_{kl}] \, [-G'_{0,kj\sigma}(i\omega_{n})]
  +
  \cdots 
\ee
or, using a matrix formulation and after elimination of the signs,
\be
  \ff G_{0}
  = 
  \ff G'_{0}
  +
  \ff G'_{0} \ff V \ff G'_{0}
  +
  \ff G'_{0} \ff V \ff G'_{0} \ff V \ff G'_{0}
  +
  \cdots
  = 
  \ff G'_{0}
  +
  \ff G'_{0} \ff V 
  (\ff G'_{0}
  +
  \ff G'_{0} \ff V \ff G'_{0}
  +
  \cdots)
  =
  \ff G'_{0}
  +
  \ff G'_{0} \ff V \ff G_{0} \: .
\ee 
In this way we have simply re-derived \refeq{cpt0} diagrammatically.
This is not yet the CPT equation as the Hubbard interaction has been disregarded. 

\subsection*{Diagram language for systems with Coulomb interaction}

\begin{figure}[t]
\includegraphics[width=0.8\textwidth]{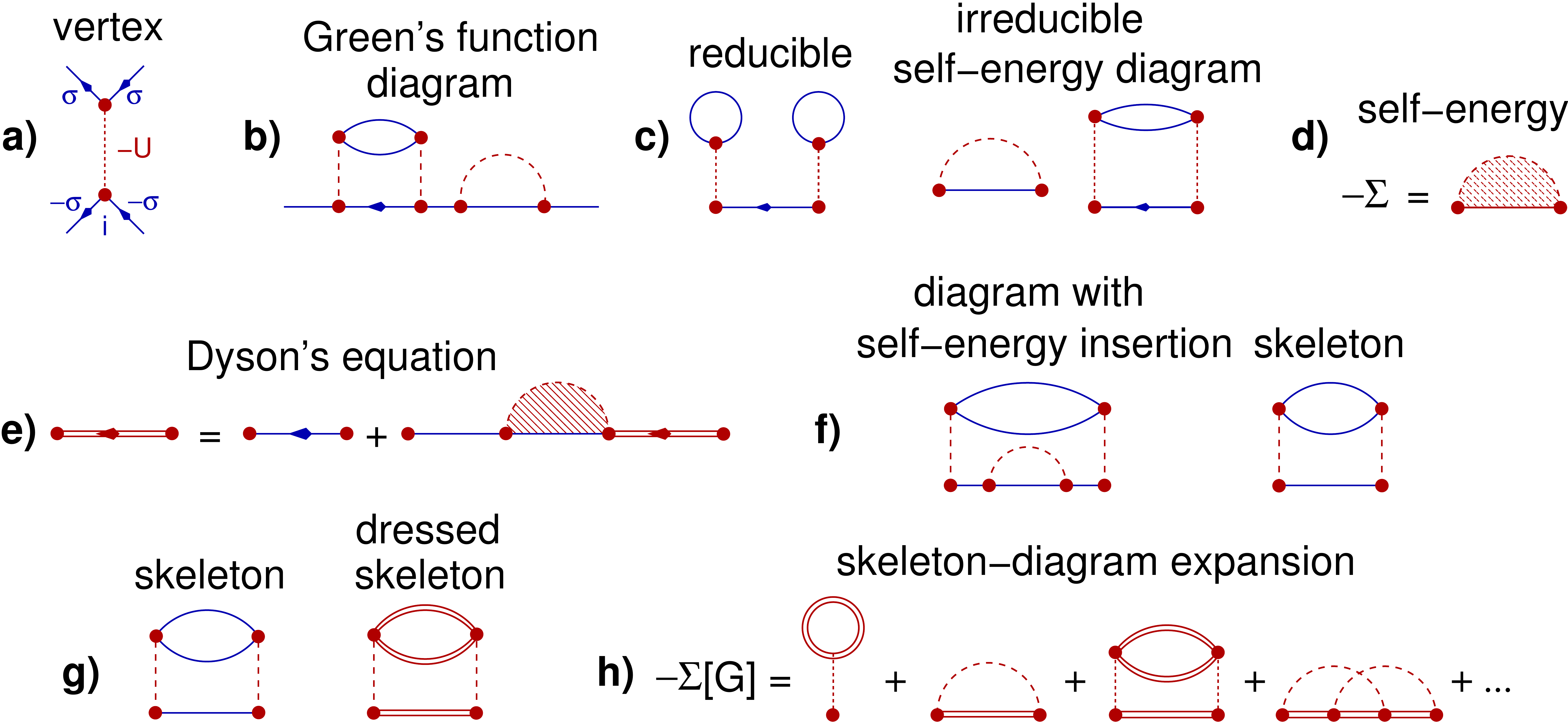}
\centering
\caption{
Diagram language for systems with Hubbard interaction. See text for discussion.
}
\label{fig:diag}
\end{figure}

Next, let us consider the system given by the Hamiltonian $H_{0}(\ff t) + H_{1}$, see Eqs.\ (\ref{eq:h0}) and (\ref{eq:h1}), and treat the Hubbard (or Coulomb) term $H_{1}$ as the interaction, as usual. 
In this case, the free propagator is given by $-\ff G_{0}$ (Fig.\ \ref{fig:scatter}a). 
To represent the interaction $-U$, we need a symbol (red dotted line) with four links, two for ``outgoing'' and two for ``incoming'' propagators (Fig.\ \ref{fig:diag}a) corresponding to the two creators and the two annihilators in the Hubbard interaction term.
Note that the interaction is local and labelled by a site index and that there is energy and spin conservation at a vertex.
A diagram contributing to the ``interacting propagator'' $-G_{ij\sigma}(i\omega_{n})$ at order $k$ consists of $2k+1$ propagators fully connecting the $k$ vertices among each other and with the two external links at the sites i and j. 
Opposed to the potential-scattering problem discussed above, there are much more diagrams at a given order $k$, namely $(2k+1)!$, one of which, for $k=3$, is shown in Fig.\ \ref{fig:scatter}b). 
The exact Green's function $\ff G(i\omega_{n})$ is obtained by summing the algebraic expressions corresponding to those diagrams and summing over all $k$. 
The detailed rules necessary for the evaluation of diagrams (see Refs.\ \cite{AGD64,FW71,NO88}) are not needed here as we do {\em not} intend to construct a diagrammatically defined approximation by summing a certain subclass of diagrams. 
While this would be the standard procedure of many-body perturbation theory, here we just want to ``speak diagrammatically''.

\index{self-energy}
\index{Dyson's equation}

One can identify so-called self-energy insertions in the diagrammatic series, i.e.\ parts of diagrams which have links to two external propagators. 
Examples are given in Fig.\ \ref{fig:diag}c where we also distinguish between reducible and irreducible self-energy insertions. 
The reducible ones can be split into two disconnected parts by removal of a single propagator line.
The self-energy $\Sigma_{ij\sigma}(i\omega_{n})$ is then defined diagrammatically as the sum over all irreducible self-energy insertions, see Fig.\ \ref{fig:diag}d.
With this we can derive Dyson's equation
\be
\ff G(i\omega_{n}) = \ff G_{0}(i\omega_{n}) +  \ff G_{0}(i\omega_{n}) \ff \Sigma(i\omega_{n}) \ff G(i\omega_{n})
\labeq{dyson}
\ee 
corresponding to Fig.\ \ref{fig:diag}e.
The double line stands for the interacting propagator $-\ff G(i\omega_{n})$.
Note that the self-energy plays the same role for the Coulomb interacting system as $\ff V$ does for the scattering problem. 

As e.g.\ the first diagram in Fig.\ \ref{fig:diag}f shows, there are irreducible self-energy diagrams which contain self-energy insertions. 
Diagrams without any self-energy insertion are called skeleton diagrams. 
Skeleton diagrams can be ``dressed'' by replacing in the diagram the free propagators with interacting propagators (double lines), see Fig.\ \ref{fig:diag}g.
It is easy to see that the self-energy is given by the sum of the skeleton diagrams only, provided that these are dressed, see Fig.\ \ref{fig:diag}h.
Therewith, the self-energy is given in terms of the interacting Green's function, $\ff \Sigma = \ff \Sigma[\ff G]$.
Only with the help of the diagram language, this very important functional relationship, called skeleton-diagram expansion, can be defined rigorously. 
If combined with Dyson's equation (\ref{eq:dyson}), it provides us with a closed equation 
\be
\ff G(i\omega_{n}) = \frac{1}{\ff G_{0}(i\omega_{n})^{-1} - \ff \Sigma[\ff G](i\omega_{n})}
\labeq{gexact}
\ee
the solution of which is given by the {\em exact} Green's function.
It is clear, however, that the functional $\ff \Sigma[\ff G]$ is extremely complicated and actually cannot be given in an explicit form -- even for the most simple models, such as the Hubbard model, and even in cases, like in the case of a small isolated Hubbard cluster, where a numerical computation of the self-energy and the Green's function is easily possible.

\subsection*{Diagrammatic derivation of the CPT}

\index{potential scattering}

Equipped with the diagrammatic language, let us come back to the central topic. 
We have $H = H_{0}(\ff t') + H_{0}(\ff V) + H_{1}$ where the reference system $H' = H_{0}(\ff t') + H_{1}$ is easily solvable since it consists of decoupled small clusters, and where $H_{0}(\ff V)$ is the inter-cluster hopping. 
Ideally, one would start from the solution of $H'$ and perform a perturbative treatment of $H_{0}(\ff V)$. 
This, however, is not possible (within the above-described standard perturbation theory) as the starting point $H'$ is an interacting system and, therefore, Wick's theorem does not apply. 
On the other hand, nothing prevents us from starting with 
$H_{0}(\ff t')$ and treating both, the inter-cluster hopping and the Hubbard interaction, $H_{0}(\ff V)$ and $H_{1}$, as the perturbation. 

\begin{figure}[t]
\includegraphics[width=0.75\textwidth]{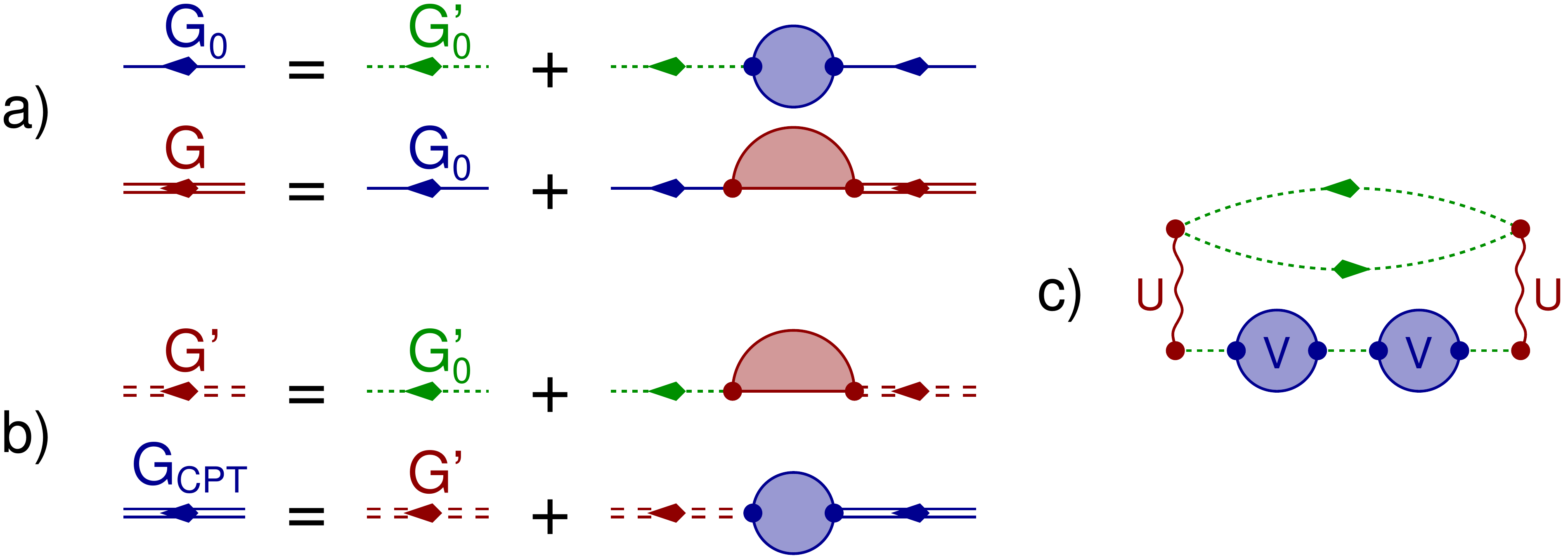}
\centering
\caption{
Diagrammatic derivation of the CPT, see text for discussion and Ref.\ \cite{BP11}.
}
\label{fig:cpt}
\end{figure}

There are two ways to do this:
(i) We start from the free ($U=V=0$) propagator $\ff G'_0$ of $H_{0}(\ff t')$ and, in a first step, sum the diagrams of all orders in $V$ but for $U=0$ -- see first line in Fig.\ \ref{fig:cpt}a. 
Merely a geometrical series must be summed which can be done exactly. 
This step has been discussed already in Sec.\ \ref{sec:scatter}. 
In a subsequent step, the resulting propagator $\ff G_0$ is dressed by taking into account the Hubbard interaction to all orders -- see second line in Fig.\ \ref{fig:cpt}a. 
This summation would yield the full Green's function $\ff G$ but obviously cannot be done in practice.

We therefore consider an alternative and reverse the order of the two steps:
First, the free ($U=V=0$) propagator $\ff G'_0$ is renormalized by the electron-electron interaction $U$ to all orders but at $V=0$ (first line in Fig.\ \ref{fig:cpt}b).
This yields the fully interacting cluster Green's function $\ff G'$. 
While, of course, $\ff G'$ cannot be computed by the extremely complicated summation of individual $U$ diagrams, it is nevertheless easily accessible via a direct (numerical) calculation if the cluster size is sufficiently small (see Sec.\ \ref{sec:green}).
In the second step, the $V=0$ propagator $\ff G'$ is renormalized due to inter-cluster potential scattering. 
Again, this is easily done by summing a geometrical series but only yields an approximation $\ff G_{\rm CPT}$ to the exact Green's function $\ff G$. 
In fact, as the second line in Fig.\ \ref{fig:cpt}b demonstrates, this is just the cluster-perturbation theory, see Eq.\ (\ref{eq:cpt}).

Note that the CPT equation (\ref{eq:cpt}) has been introduced in an {\em ad hoc} way. 
Contrary, the diagram approach allows to understand the CPT as an approximation that is given by summing a certain subclass of diagrams. 
Fig.\ \ref{fig:cpt}c displays a low-order self-energy diagram which is neglected in this summation. 
This clearly shows that the CPT cannot be correct and suggests two different routes for improvement, namely (i) taking into account missing diagrams and (ii) using the freedom in the CPT construction to optimize the starting point. 
The first idea is related to the attempt to perform a systematic perturbational expansion around the disconnected-cluster limit and is notoriously complicated (as Wick's theorem does not apply) \cite{RKL08}.

Another {\em ad hoc} way to derive the CPT follows the idea of the so-called Hubbard-I approximation \cite{Hub63}: 
The main idea is to employ the Dyson equation (\ref{eq:dyson}) {\em of the reference system} to compute the reference system's self-energy,
\be
  \ff \Sigma'(i\omega_{n}) = \ff G'_{0}(i\omega_{n})^{-1} - \ff G'(i\omega_{n})^{-1} \: ,
\labeq{sigmacpt}  
\ee
and to consider this as an approximation for the self-energy of the original system: $\ff \Sigma(i\omega_{n}) \approx \ff \Sigma'(i \omega_{n})$. 
The motivation for this step is that the self-energy, opposed to the Green's function, is a much more local object as is well known at least for the weak-coupling regime from standard perturbation theory \cite{SC90,SC91,PN97c} and by the fact that the self-energy becomes purely local in the limit of lattices with infinite spatial dimensions \cite{MV89,MH89b}. 
Furthermore, the idea is reminiscent of dynamical mean-field theory (DMFT) \cite{MV89,GKKR96,KV04} where the self-energy of an impurity model approximates the self-energy of the lattice model. 
Using \refeq{sigmacpt} in Dyson's equation for the original model, we find:
\be
  \ff G(i\omega_{n})
  =
  \frac{1}{
  \ff G_{0}(i\omega_{n})^{-1} - (\ff G'_{0}(i\omega_{n})^{-1} - \ff G'(i\omega_{n})^{-1}) 
  }
  =
  \frac{1}{
  \ff G'(i\omega_{n})^{-1} - \ff V
  } \; ,
\ee
i.e.\ the CPT equation (\ref{eq:cpt}) is recovered.
We note in passing that the Hubbard-I approach is obtained is for $L_{\rm c}=1$ and with some {\em ad hoc} element of self-consistency \cite{Hub63}.

This way to construct the CPT suggests to use, rather than the Ritz principle, a variational principle of the form
\be
  \frac{\delta \Omega[\ff \Sigma]}{\delta \ff \Sigma(i\omega_{n})} = 0 \; ,
\ee
where the trial self-energy is taken from the reference system and varied by varying the parameters of the reference system. 
Ideally, this ``self-energy-functional approach'' should also cure the different defects of the CPT, i.e.\ besides
the arbitrariness of the CPT construction, 
the non-self-consistent nature of the approach and its inability to describe spontaneous symmetry breaking as well as different thermodynamical inconsistencies that show up in the computation of a thermodynamical potential from the Green's function \cite{AGD64,FW71,NO88}.
Furthermore, one may ask whether both, the CPT and the DMFT, can be understood in single unifying theoretical framework.

\section{Self-energy functional theory}

\index{self-energy functional theory}

\subsection*{Luttinger-Ward generating functional}

\index{Luttinger-Ward functional}

\begin{figure}[t]
  \centerline{\includegraphics[width=0.45\textwidth]{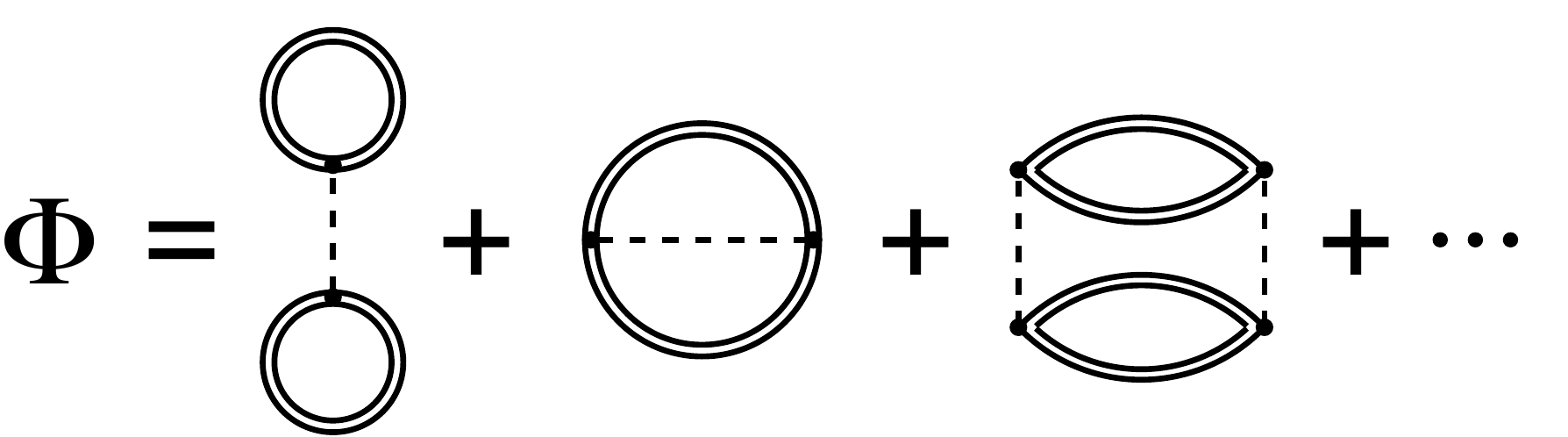}}
\caption{
Diagrammatic construction of the Luttinger-Ward functional $\Phi[\ff G]$. 
Double lines stand for fully interacting propagators, dashed lines for the Hubbard interaction.
}
\label{fig:lw}
\end{figure}

The construction of a variational principle based on the self-energy is in fact possible with the help of the so-called Luttinger-Ward functional \cite{LW60}. 
This is a scalar functional $\Phi$ of the Green's function $\ff G$ which has originally been defined by all-order perturbation theory
(a construction that uses the path integral can be found in Ref.\ \cite{Pot06b}). 
Namely, $\Phi[\ff G]$ is defined as the sum of all closed, connected and fully dressed skeleton diagrams of any order $k$.
Fig.\ \ref{fig:lw} shows the lowest-order diagrams. 
Closed diagrams without links to external propagators are diagrams contributing to the partition function, see \refeq{p1}. 
The Luttinger-Ward series is given by dressed skeleton diagrams, i.e.\ diagrams without self-energy insertions where the free propagators are replaced by the fully interacting ones. 
One easily verifies that, due to dressing of the diagrams, some diagrams in the expansion of $Z/Z_{0}$ are counted twice or more. 
This is done on purpose. 
Namely, the most important property of the Luttinger-Ward functional constructed in this way is that its functional derivative just yields the skeleton-diagram expansion of the self-energy:
\be
  \frac{\delta {\Phi}[\ff G]}{\delta \ff G(i\omega_{n})} 
  = 
  \frac{1}{\beta}
  {\ff \Sigma}[\ff G](i\omega_{n}) \: .
\label{eq:der}
\ee 
This can be verified, diagram by diagram: 
The functional derivative of a dressed skeleton just corresponds to the removal of a dressed propagator and results in a dressed skeleton diagram with two links for external propagators which contributes to the self-energy.
When carefully taking into account the coefficients of the two different expansions, \refeq{p1} and \refeq{p2}, one easily derives Eq.\ (\ref{eq:der}).
The equation is remarkable as it shows that the different components of the self-energy $\Sigma_{ij\sigma}(i\omega)$ can be obtained from the {\em scalar} functional. 
In fact, the existence of the Luttinger-Ward functional can also be proven by verifying a vanishing-curl condition as has been done by Baym and Kadanoff \cite{BK61,Bay62}.

The value $\Phi$ of the Luttinger-Ward functional has no direct physical meaning. 
Summing all closed diagrams (not only connected skeletons) yields the partition function $Z/Z_{0}$ -- by construction. 
Summing connected diagrams only, yields $\ln Z$ as is known from the linked-cluster theorem \cite{LW60,AGD64}.
The sum of dressed connected skeletons, however, cannot provide the grand potential $\propto \ln Z$ because of the above-mentioned double counting.

\subsection*{Self-energy functional}

We will make use of $\Phi[\ff G]$ by defining the following functional of the self-energy:
\be
  {\Omega}[\ff \Sigma] = 
  \Tr \ln \frac{1}{\ff G_{0}^{-1} - \ff \Sigma}
  + 
  {\Phi}[{\ff G}[\ff \Sigma]] 
  - 
  \Tr (\ff \Sigma {\ff G}[\ff \Sigma]) 
  \: .
\labeq{sef}
\ee
Here, the frequency dependencies are suppressed in the notations and 
\be
\Tr \, \ff A \equiv \frac{1}{\beta} \sum_n \sum_{i\sigma} e^{i\omega_n0^+} A_{ii\sigma}(i\omega_n)
\ee
is used where $0^+$ is a positive infinitesimal.
Furthermore, ${\ff G}[{\ff \Sigma}]$ is the inverse of the functional $\ff \Sigma[\ff G]$, i.e.\ ${\ff G}[{\ff \Sigma}[\ff G]] = \ff G$. 
We assume that this inverse of the skeleton-diagram expansion of the self-energy exists at least {\em locally}. 
The second part of the self-energy functional,
\be
  {F}[\ff \Sigma] 
  \equiv
  {\Phi}[{\ff G}[\ff \Sigma]] 
  - 
  \Tr (\ff \Sigma {\ff G}[\ff \Sigma]) \: ,
\ee
is just the Legendre transform of the Luttinger-Ward functional. 
With ${\ff \Sigma}[{\ff G}[\ff \Sigma]] = \ff \Sigma$ and \refeq{der} we immediately have
\be
\frac{\delta {F}[\ff \Sigma]}{\delta \ff \Sigma} 
=
-
\frac{1}{\beta}
{\ff G}[\ff \Sigma] \: .
\ee
Therewith, we can also calculate the functional derivative of $\Omega[\ff \Sigma]$:
\be
  \frac{\delta {\Omega}[\ff \Sigma]}
  {\delta \ff \Sigma} 
  = 
  \frac{1}{\beta}
  \left(
  \frac{1}{\ff G_{0}^{-1} - \ff \Sigma} - 
  {\ff G}[\ff \Sigma] 
  \right)
  \: .
\ee
The equation 
\be
  {\ff G}[\ff \Sigma] = \frac{1}{\ff G_{0}^{-1} - \ff \Sigma}
\label{eq:sig}
\ee
is a (highly non-linear) conditional equation for the self-energy of the system $H = H_0(\ff t) + H_{1}$.
Inserting $\ff \Sigma = \ff \Sigma[\ff G]$ shows that it is (locally) equivalent to \refeq{gexact}.
It is satisfied by the exact self-energy of the system.
Therefore, solving Eq.\ (\ref{eq:sig}) is equivalent to a search for the stationary point of the self-energy functional:
\be
  \frac{\delta {\Omega}[\ff \Sigma]} {\delta \ff \Sigma} = 0 \; .
\label{eq:var}
\ee
This represents the dynamical variational principle we have been looking for.
The exact self-energy of the system makes the self-energy functional $\Omega[\ff \Sigma]$, \refeq{sef}, stationary. 

\index{dynamical variational principle}

The definition of the self-energy functional given with \refeq{sef} is a formal one only. 
The argument of the $\ln$ is not dimensionless and furthermore, since $\ff G(i\omega_{n}) \propto 1/\omega_{n} \propto 1/(2n+1)$ for large $n$, the sum over the Matsubara frequencies, $\sum_{n} \ln (2n+1)$, does not converge.
This problem can be solved, however, by replacing $\Omega[\ff \Sigma] \mapsto \Omega[\ff \Sigma] - \Tr \ln \ff G_{\rm reg}$ with $G^{-1}_{{\rm reg},ij\sigma}(i\omega_{n}) = \delta_{ij}(i\omega_{n} - \varepsilon_{\rm reg})$ and taking the limit $\varepsilon_{\rm reg} \to \infty$ {\em after} all calculations are done. 
As the constant $\Tr \ln \ff G_{\rm reg}$ does not depend on $\ff \Sigma$, the variational principle is unaffected but now the Matsubara sum over both logarithms is well defined and convergent.
One can show \cite{LW60,AGD64,Pot11} that, if evaluated at the physical (exact) self-energy, the regularized $\Omega[\ff \Sigma] - \Tr \ln \ff G_{\rm reg}$ is just the grand potential of the system.
This provides us with a physical interpretation of the self-energy functional.  
In the following this regularization is always implicit.

As a remark, we note that at $U=0$ the self-energy functional reduces to the expression $\Omega_{0} \equiv \Tr \ln \ff G_{0}$ as becomes obvious from the diagrammatic definition of $\Phi[\ff G]$ and of $\ff \Sigma$ since there are simply no diagrams left at zero-th order in the interaction strength:
\be
  \Phi[\ff G] \equiv 0 \: , \; \ff \Sigma(i\omega_{n}) = 0 \qquad \mbox{for $U=0$} \: .
\ee
If regularized properly, $\Omega_{0} \mapsto \Omega_{0} - \Tr \ln \ff G_{\rm reg}$, this exactly yields the grand potential of the non-interacting system.

\subsection*{Evaluation of the self-energy functional} 

The diagrammatic definition of the Luttinger-Ward functional (Fig.\ \ref{fig:lw}) uncovers another remarkable property:
Since any diagram contributing to $\Phi$ consists of vertices and dressed propagators only, the functional relation ${\Phi}[\cdots]$ is completely determined by the interaction $U$ but does not depend on $\ff t$. 
Clearly, this ``universality'' then also holds for its Legendre transform $F[\ff \Sigma]$: 
Two systems (at the same chemical potential $\mu$ and inverse temperature $\beta$) with the same interaction $H_{1}$ but different one-particle parameters $\ff t$ and $\ff t'$ are described by the same functional $F[\ff \Sigma]$. 
Contrary, the first part of the self-energy functional, 
\be
  {\Omega}[\ff \Sigma] 
  = 
  \Tr \ln \frac{1}{\ff G_{0}^{-1} - \ff \Sigma}
  + 
  {F}[\ff \Sigma] \: 
\labeq{sfp}
\ee
does depend on the hopping, namely via $\ff G_{0}^{-1}(i\omega_{n}) = i\omega_{n} + \mu - \ff t$, but not on the interaction strength $U$.

\begin{figure}[t]
  \centerline{\includegraphics[width=0.4\textwidth]{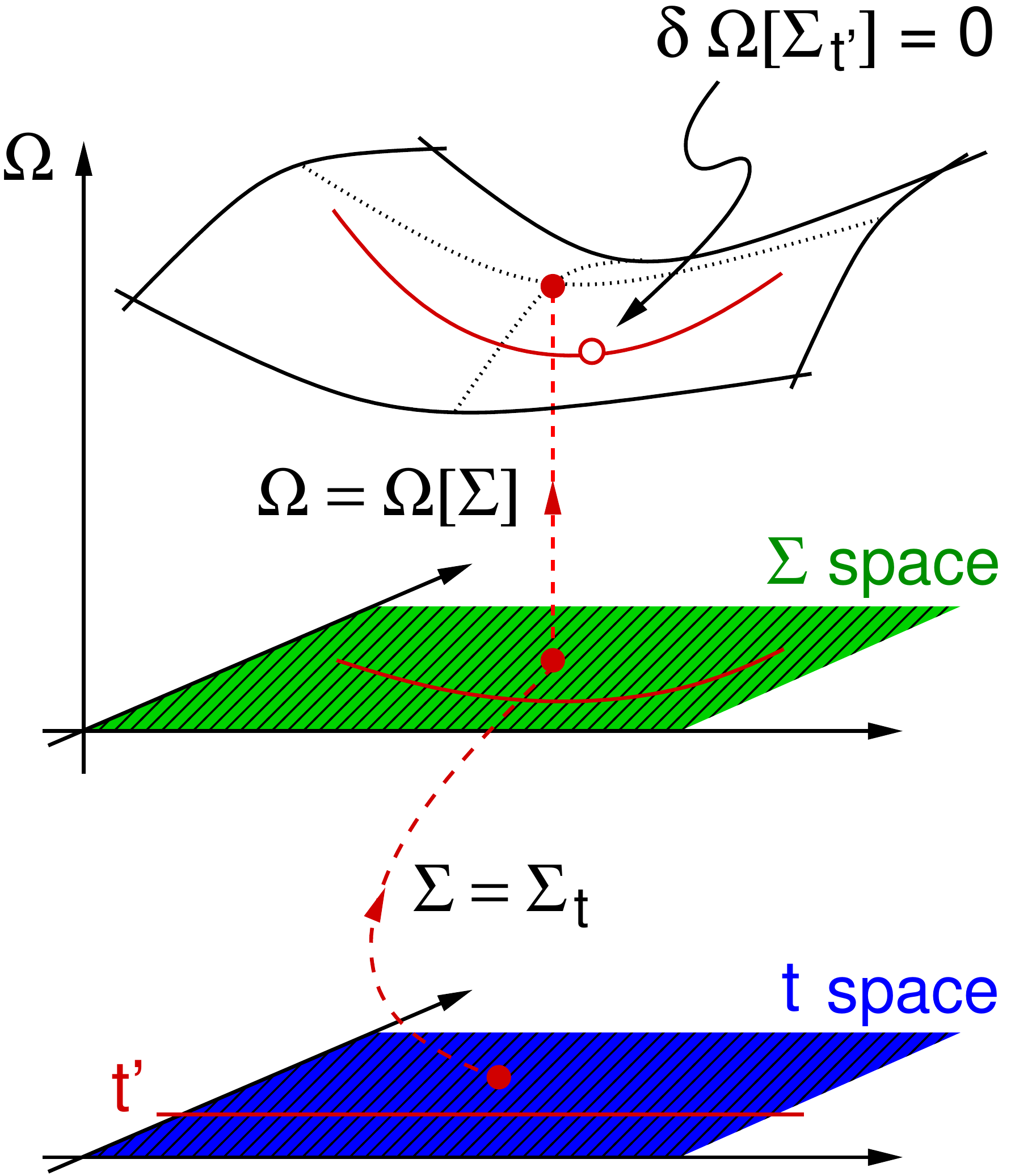}}
\caption{
Construction of consistent approximations within the self-energy-functional theory.
The grand potential is considered as a functional of the self-energy which is stationary at the physical (exact) self-energy $\ff \Sigma_{\ff t}$ (filled red circles).
The functional dependence of $\Omega [\ff \Sigma]$ is not accessible on the entire space of self-energies ($\Sigma$ space). 
However, $\Omega [\ff \Sigma]$ can be evaluated exactly on a restricted subspace of trial self-energies $\ff \Sigma_{\ff t'}$ parametrized by a subset of one-particle parameters $\ff t'$ (solid red lines). These $\ff t'$ define an exactly solvable ``reference system'', i.e.\ a manifold of systems with the same interaction part but a modified one-particle part given by $\ff t'$. 
Typically, the reference system consists of a set of decoupled clusters.
A self-energy at which the grand potential is stationary on this sub-manifold represents the approximate self-energy of the original system and the grand potential at this self-energy represents the approximate grand potential (open circle).
}
\label{fig:sft}
\end{figure}

The universality property of $F[\ff \Sigma]$ is more important, however, as this functional is basically unknown.
The central idea of self-energy-functional theory is to compare the self-energy functional of two systems, the original system with $H=H_{0}(t) + H_{1}$ and the reference system with $H'=H_{0}(\ff t') + H_{1}$, i.e.\ \refeq{sfp} and
\be
  {\Omega'}[\ff \Sigma] = 
  \Tr \ln \frac{1}{{\ff G'_{0}}^{-1} - \ff \Sigma}
  + F[\ff \Sigma] \: .
\labeq{sfpref}
\ee
Due to its universality, one can eliminate the unknown functional $F[\ff \Sigma]$ by combining both equations:
\be
  {\Omega}[\ff \Sigma] 
  = 
  {\Omega'}[\ff \Sigma] 
  + 
  \Tr \ln \frac{1}{\ff G_{0}^{-1} - \ff \Sigma}
  - 
  \Tr \ln \frac{1}{{\ff G'_{0}}^{-1} - \ff \Sigma} 
  \: .
  \labeq{sfp1}
\ee
This equation is still exact.
Since the functional dependence of ${\Omega'}[\ff \Sigma]$ is unknown, also in the case of a simple reference system with decoupled clusters, it appears that this step amounts to a mere shift of the problem only.
The great advantage of \refeq{sfp1} becomes manifest, however, when inserting the exact self-energy of the reference system $\ff \Sigma_{\ff t'}(i\omega_{n})$ as a trial self-energy. 
(We use the notation $\ff \Sigma_{\ff t'}$ for the exact self-energy of the system with hopping parameters $\ff t'$ and interaction $H_{1}$.)
Namely, the first term on the r.h.s.\ of \refeq{sfp1} then just reduces to the grand potential of the reference system $\Omega'$ which can be computed easily if, as we assume, the reference is amenable to an exact numerical solution. 
The same holds for the second and the third term. 
We find:
\be
  {\Omega}[\ff \Sigma_{\ff t'}] 
  = 
  {\Omega}'
  + 
  \Tr \ln \frac{1}{\ff G_{0}^{-1} - \ff \Sigma_{\ff t'}}
  - 
  \Tr \ln \frac{1}{{\ff G'_{0}}^{-1} - \ff \Sigma_{\ff t'}}
  \:  .
\labeq{ocalc}	
\ee
This is a remarkable result.
It shows that an {\em exact} evaluation of the self-energy functional of a non-trivial interacting system is possible, at least for trial self-energies which are taken from an exactly solvable reference system with the same interaction part  (see Fig.\ \ref{fig:sft}).

\subsection*{The variational cluster approximation}

\index{variational cluster approximation}

We recall that the cluster-perturbation theory approximates the self-energy of the original lattice-fermion model by the self-energy of a reference system of disconnected clusters. 
As one may choose the intra-cluster parameters of the reference system different from the corresponding parameters of the original system, there is a certain arbitrariness in the CPT construction. 
Usually, one simply assumes that e.g.\ the intra-cluster nearest-neighbor hopping of the reference system is the same as the physical hopping. 
There are, however, good reasons not to do so. 
One example are symmetry breaking Weiss fields, as already mentioned above.
Another one becomes obvious from Fig.\ \ref{fig:cpt}c, where the CPT is seen to neglect the effect of the scattering at the inter-cluster potential on the self-energy.
Therefore, an enhanced {\em intra}-cluster hopping could, at least partially, compensate for the missing feedback of the {\em inter}-cluster hopping on the approximate self-energy.

With the self-energy-functional framework at hand, we can now remove the arbitrariness of the CPT approach and determine the ``optimal'' self-energy from \refeq{ocalc}.
This optimal self-energy is the exact self-energy of an optimized reference system that is specified by a set of one-particle (intra-cluster) parameters $\ff t'$. 
Note that to derive \refeq{ocalc} it was necessary to assume that the interaction part $H_{1}$ of the reference system cannot be optimized and must be the same as the interaction of the original system.
Therefore, the role of the reference system is to generate a manifold of trial self-energies $\ff \Sigma_{\ff t'}$ which are parameterized by the one-particle parameters $\ff t'$.
As the self-energy functional \refeq{sfp} can be evaluated {\em exactly} on this manifold via \refeq{ocalc}, the optimal self-energy $\ff \Sigma_{\ff t'_{\rm opt}}$is given as the solution of the SFT Euler equation
\be
  \frac{\partial \Omega[\ff \Sigma_{\ff t'}]}{\partial \ff t'} \Bigg|_{\ff t' = \ff t'_{\rm opt}} = 0 \: .
\labeq{euler}
\ee
For a cluster reference system, this equation constitutes the variational cluster approximation (VCA). 

The VCA represents an approximation as it provides the stationary point of the self-energy functional on a restricted manifold of trial self-energies only rather than on the entire ``self-energy space'' (see Fig.\ \ref{fig:sft}).
The latter could be defined as the set of the self-energies of all models with the interaction part fixed at $H_{1}$ but with an completely arbitrary on-particle part (that also may connect the clusters). 
This space, of course, contains the exact self-energy $\ff \Sigma_{\ff t}$ of our lattice model $H_{0}(\ff t) + H_{1}$ while the optimized VCA self-energy $\ff \Sigma_{\ff t'_{\rm opt}}$ is constrained to the manifold of cluster trial self-energies.
Approximations generated in this way have a number of advantageous properties: 
First of all, although we have employed the language of perturbation theory, the VCA is non-perturbative. 
Formally, the diagram series has not been cut at any level, no subclass of diagrams is neglected etc.
The approximation rather results from a restricted domain of a self-energy functional. 
Second, the VCA is an internally consistent approximation in the sense that all observables derive from an explicit (though approximate) expression for a thermodynamical potential, namely from the self-energy functional evaluated at the optimal self-energy $\Omega[\ff \Sigma_{\ff t'_{\rm opt}}]$.
Third, the VCA can be improved in a systematic way by increasing the cluster size $L_{\rm c}$ as one has to approach the exact solution for $L_{\rm c}\to \infty$. 
Here, the reference system is basically identical with the original system, and $\ff t$ is basically {\em within} the space of the variational parameters $\ff t'$.
One may interpret $1/L_{\rm c}$ as the ``small parameter'' that ``controls'' the quality of the approximation.
It is clear, however, that the numerical effort to solve \refeq{euler} also increases with $L_{\rm c}$. 
This increase is even exponential if an exact-diagonalization ``solver'' is used to compute the cluster self-energy, Green's function and grand potential that enter \refeq{ocalc}.
Unfortunately, one cannot {\em a priori} give a ``distance'' by which the optimal VCA self-energy differs from the exact self-energy. 
In practice, the quality of the approximation must therefore be controlled by comparing the results obtained for different cluster sizes $L_{\rm c}$. 
For small clusters, also the cluster geometry and the imposed cluster boundary conditions matter, and must be checked.

Although the VCA derives from a general variational principle, it is not ``variational'' in the sense that the approximate VCA grand potential $\Omega[\ff \Sigma_{\ff t'_{\rm opt}}]$ must always be larger than the exact grand potential $\Omega$. 
Opposed to the Ritz principle and the state functional $E[\ket{\Psi}]$ or, at finite temperatures, the density-matrix functional $\Omega[\rho]$, the self-energy functional $\Omega[\ff \Sigma]$ is not convex and hence there is no reason to assume that the VCA provides an upper bound to $\Omega$.

Concluding, the VCA must be seen as a cluster mean-field approximation that focusses on one-particle correlations, the one-particle excitation spectrum (e.g.\ photoemission) and thermodynamics. 
It treats short-range one-particle correlations within the cluster in an explicit way while inter-cluster one-particle correlations are accounted for via Dyson's equation. 
The feedback of local and intra-cluster {\em two}-particle (and even higher) correlations on the one-particle self-energy is explicitly and non-perturbatively taken into account while the feedback of non-local two-particle, e.g.\ magnetic, correlations on the one-particle spectrum is neglected altogether. 
This is typical for cluster mean-field theories \cite{MJPH05} and should be kept in mind when studying e.g.\ systems close to a second-order phase transition where non-local correlations play an important role.

\section{Implementation of the variational cluster approximation}

\subsection*{$Q$-matrices}

The bottleneck of a practical VCA calculation consists in the computation of the Green's function of the reference system.
Using an exact-diagonalization technique, the Green's function for an individual cluster can be obtained in its Lehmann representation, see \refeq{lehmann}. 
Let $\alpha=(i,\sigma)$ be an index referring to the elements of the localized orbitals forming an orthonormal basis of the one-particle Hilbert space.
Therewith the elements of the cluster Green's function can be written in the form
\begin{equation}
  G'_{\alpha\beta}(\omega) = \sum_m
  Q'_{\alpha m} \frac{1}{\omega - \omega'_m} {Q'}^\dagger_{m\beta} \; . 
\end{equation}
Here, $m=(r,s)$ refers to a single-particle excitation between two energy eigenstates $|s\rangle$ and $|r\rangle$ of the (grand-canonical) Hamiltonian of the reference system $H' - \mu N$, and $\omega'_m = E'_r - E'_s$ is the excitation energy.
$Q'_{\alpha m}$ are the elements of the so-called $Q'$-matrix \cite{AAPH06b} which is a rectangular matrix with a small number of rows but a large number of columns (dimension of the one-particle Hilbert space $\times$ number of many-body excitations):
\be
  Q'_{\alpha m} = \langle r | c_{\alpha} | s \rangle \sqrt{\frac{\exp(-\beta E'_{r}) + \exp(-\beta E'_{s})}{Z'}} \; , \quad  Z' = \sum_{r} e^{-\beta E'_{r}} \: ,
\ee
as is readily read off from the Lehmann representation \refeq{lehmann}. 
One also verifies that $\ff Q' \ff {Q'}^\dagger = \ff 1 \ne \ff {Q'}^\dagger \ff Q'$.
Using the $Q'$-matrix, we can write the reference system's Green's function in a compact form as
\be
\ff G'(\omega) = \ff Q' \frac{1}{\omega - \ff \Lambda'} \ff {Q'}^{\dagger} \: ,
\labeq{gpq}
\ee
where $\ff \Lambda'$ is the diagonal matrix with elements $\Lambda'_{mn} = \omega^\prime_m \delta_{mn}$.
With $\ff V = \ff t - \ff t^\prime$, the Green's function of the original system is obtained as:
\ba
\ff G(\omega) 
&=&
\frac{1}{\ff G'(\omega)^{-1} - \ff V}
=
\ff G'(\omega) + \ff G'(\omega) \ff V \ff G'(\omega) + \cdots
\nonumber \\
&=&
\ff Q' \left(
\frac{1}{\omega - \ff \Lambda'}
+
\frac{1}{\omega - \ff \Lambda'}
\ff {Q'}^{\dagger} \ff V \ff Q'
\frac{1}{\omega - \ff \Lambda'}
+
\cdots
\right)
\ff {Q'}^{\dagger}
=
\ff Q' 
\frac{1}{\omega - \ff M}
\ff {Q'}^{\dagger} \: ,
\ea
where $\ff M = \ff \Lambda' + \ff {Q'}^\dagger \ff V \ff {Q'}$ is a (large) square Hermitian matrix which can be diagonalized by a unitary transformation, $\ff M=\ff S \ff \Lambda \ff S^{\dagger}$. 
Here, $\Lambda_{mn} = \omega_m \delta_{mn}$ with the poles $\omega_{m}$ of $\ff G(\omega)$. 
We find
\be
\ff G(\omega) 
=
\ff Q \frac{1}{\omega - \ff \Lambda} \ff {Q}^{\dagger} 
\labeq{gq}
\ee
with $\ff Q = \ff Q' \ff S$.
The representations \refeq{gpq} and \refeq{gq} are particularly useful to evaluate the self-energy functional \refeq{ocalc} in practice.
The trace $\Tr$ contains a Matsubara-frequency summation which can be carried out analytically \cite{Pot03b} such that one is left with a simple algebraic expression \cite{AAPH06b}, 
\be
  \Tr \ln \frac{1}{\ff G_{0}^{-1} - \ff \Sigma_{\ff t'}}
  - 
  \Tr \ln \frac{1}{{\ff G'_{0}}^{-1} - \ff \Sigma_{\ff t'}}
  =
  - \sum_{m} \frac{1}{\beta} \ln (1+e^{-\beta \omega_{m}})
  + \sum_{m} \frac{1}{\beta} \ln (1+e^{-\beta \omega'_{m}}) \; ,
\labeq{trlng}
\ee
which involves the poles of $\ff G(\omega)$ and $\ff G'(\omega)$ only.
Finally, the grand potential of the reference system in \refeq{ocalc} is easily computed as $\Omega' = - (1/\beta) \ln \sum_{r} e^{-\beta E'_{r}}$.

\subsection*{Recipe for practical calculations}

A typical VCA calculation is carried out as follows:
\bi
\item
Construct a reference system by tiling the original lattice into identical clusters. 
\item
Choose a set of one-particle parameters $\ff t'$ of the reference system and compute $\ff V = \ff t - \ff t'$.
\item
Solve the problem for the reference system ($U$ is fixed), i.e.\ compute the Green's function $\ff G'$ and find the poles $\omega'_{m}$ and the $Q'$-matrix.
\item
Get the poles $\omega_{m}$ of the approximate Green's function of the original system by diagonalization of the matrix $\ff M = \ff \Lambda' + {\ff Q'}^\dagger \ff V \ff Q$.
\item 
Calculate the value of the SFT grand potential via \refeq{ocalc} and \refeq{trlng} and by calculating the grand potential of the reference system $\Omega'$ from the eigenvalues of $H'$.
\item
Iterate this scheme for different $\ff t'$, such that one can solve 
\begin{equation}
  \frac{\partial \Omega[\ff \Sigma_{\ff t'}]}{\partial \ff t'} \Bigg|_{\ff t' = \ff t'_{\rm opt}} \stackrel{!}{=} 0 \: 
\end{equation}
for $\ff t'_{\rm opt}$.
\item
Evaluate observables, such as $\Omega[\ff \Sigma_{\ff t'_{\rm opt}}]$, $\ff G(\omega)$ and static expectation values derived from the SFT grand potential by differentiation, at the stationary point $\ff t'_{\rm opt}$.
\item
Redo the calculations for different parameters of the {\em original} system, e.g.\ a different $U$, filling or $\beta$ to scan the interesting parameter space.
\ei

\subsection*{Tips and tricks}

For a given topology of the reference system, i.e.\ for a given cluster geometry, one may in principle consider all one-particle particle parameters $\ff t'$ as variational parameters. 
However, besides an exponentially increasing Hilbert-space dimension, a larger cluster also implies an increasing numerical complexity for the search of the stationary point since $\Omega[\ff \Sigma_{\ff t'}]$ is a function of a multi-component variable $\ff t'$. 
It is therefore advisable to restrict the search to a small number of physically important parameters. 
In most cases, a few variational parameters suggest themselves. 

An overall shift $\Delta \varepsilon'$ of the on-site energies in the cluster (like the chemical potential), $t'_{ii} \mapsto t'_{ii} + \Delta \varepsilon'$, should be among the variational parameters to ensure thermodynamical consistency with respect to the total particle number as has been pointed out in Ref.\ \cite{AAPH06a}. 
This ensures that both ways to compute the total particle number, 
$\langle N \rangle = - \partial \Omega / \partial \mu$ and  
$\langle N \rangle = \sum_{i\sigma} \int dz \, A_{ii\sigma}(z) / (e^{\beta z} +1)$,
must yield the same result.
Analogously, in case of a (ferro- or antiferro-) magnetic system, one should include a (homogeneous or staggered) Weiss field $B'$ in the set of variational parameters.
For a paramagnetic system and for a system with manifest particle-hole symmetry, however, symmetry considerations {\em a priori} fix those variational parameters to $B'=0$ and $\Delta \varepsilon' = 0$, i.e.\ $t'_{ii} = t_{ii}$. This can also be verified by a practical VCA calculation.

For the setup of self-energy-functional theory it is inevitable that the original and the reference system have the same interaction $H_{1}$. 
Contrary, the one-particle part of the reference system can be designed at will. 
One very interesting option in this context is to add additional fictitious sites to the cluster. 
These ``bath sites'' have to be non-interacting ($U=0$) contrary to the ``correlated sites'' ($U>0$) which correspond to the physical sites (with the same $U$) of the original system. 
Adding the bath sites does not change the interaction part $H_{1}$ of the Hamiltonian and therefore leave the Luttinger-Ward functional as well as its Legendre transform $F[\ff \Sigma]$ unaffected.
Bath sites can be coupled via one-particle hopping terms to the correlated sites in the cluster. 
This construction has the appealing advantage to increase the space of variational parameters and thereby to improve the quality of the approximation {\em locally}. 
Adding bath sites and optimizing the additional associated parameters will improve the description of local temporal correlations while increasing the cluster improves the theory with respect to non-local spatial correlations. 

If one decides to consider a reference system with bath sites, it is advisable to formally include the same bath sites also in the original system. 
Here, of course, they are completely decoupled from the correlated sites (the respective parts of the hopping matrix $\ff t$ have to be set to zero) such that all physical quantities remain unchanged.
The advantage of this trick is that $\ff t$ and $\ff t'$ have the same matrix dimension, and that the Hamiltonians $H$ and $H'$ operate on the same Hilbert space.
The ``inter-cluster hopping'' $\ff V = \ff t - \ff t'$ includes the hopping terms between correlated and bath sites in the reference system only.

Rather than employing the above-mentioned $Q$-matrix technique, one may also perform the traces in \refeq{ocalc}, i.e.\ the trace of the spatial and orbital degrees of freedom and the implicit Matsubara-frequency summation numerically. 
This is recommendable if the dimension of $\ff M$, given by the number of poles of $\ff G^\prime$ with non-vanishing spectral weight, becomes too large.

If a full diagonalization of the cluster problem is not feasible and Krylov-space methods shall be applied, one has to make sure that the different elements $G'_{ij\sigma}(\omega)$ have the same set of poles: $\omega'_m$ should be independent of $i,j$.
This can be achieved by the band Lanczos method \cite{Fre00}. 
The dimension of the matrix $\ff M$ is given by the number of iteration steps in the Lanczos procedure. 
Typically, about 100 steps are sufficient for reasonably well converged results.
This should be checked regularly.

The SFT grand potential may exhibit more than a single stationary point. 
A minimal grand potential among the grand potentials at the different stationary points distinguishes the thermodynamically stable phase in most situations \cite{Pot06a}. 
Often, the occurrence of several stationary points is welcome from a physical point of view. 
For example, scanning a physical parameter, e.g.\ $U$, a second-order magnetic phase transition is characterized by a bifurcation of a non-magnetic solution into a non-magnetic and a magnetic one (or even more magnetic ones). 

There are different numerical strategies to determine a stationary point of the self-energy functional, see Ref.\ \cite{PTVF07} for examples.
If there is only a single variational parameter to be optimized, iterative bracketing of maxima and minima can be employed efficiently.
In the case of more than one variational parameter, the SFT grand potential usually exhibits saddle point rather than a minimum or maximum. 
A strategy that has been found to be useful for two or three parameters, is to assume (and verify) a certain characteristic of the saddle point and to 
apply iterated one-dimensional optimizations.
The downhill simplex method can be used for higher-dimensional parameter spaces to find the local minima of $|\partial \Omega[\ff \Sigma_{\ff t'}] / \partial \ff t'|^2$. 
If there is more than one, only those must be retained for which $\Omega[\ff \Sigma(\ff t')]$ has a vanishing gradient.

\section{Selected results}

The VCA is not restricted to the single-band Hubbard model but has also been applied to a variety of multi-orbital systems.
The necessary generalization to the multi-orbital case is straightforward.
In this way the VCA has contributed to the study of the correlated electronic structure of real materials
such as NiO \cite{Ede07}, CoO and MnO \cite{Ede08}, CrO$_2$ \cite{CAA+07}, LaCoO$_3$ \cite{Ede09}, TiOCl \cite{ASV+09}, TiN \cite{ACA09}, and NiMnSb \cite{ACA+10}.
Here, however, we will focus on the single-band model and discuss a few and very simple examples to illustrate the theory. 

\subsection*{One-dimensional Hubbard model}

\index{one-dimensional Hubbard model}

\begin{figure}[t]
\centering
\includegraphics[width=0.99\textwidth]{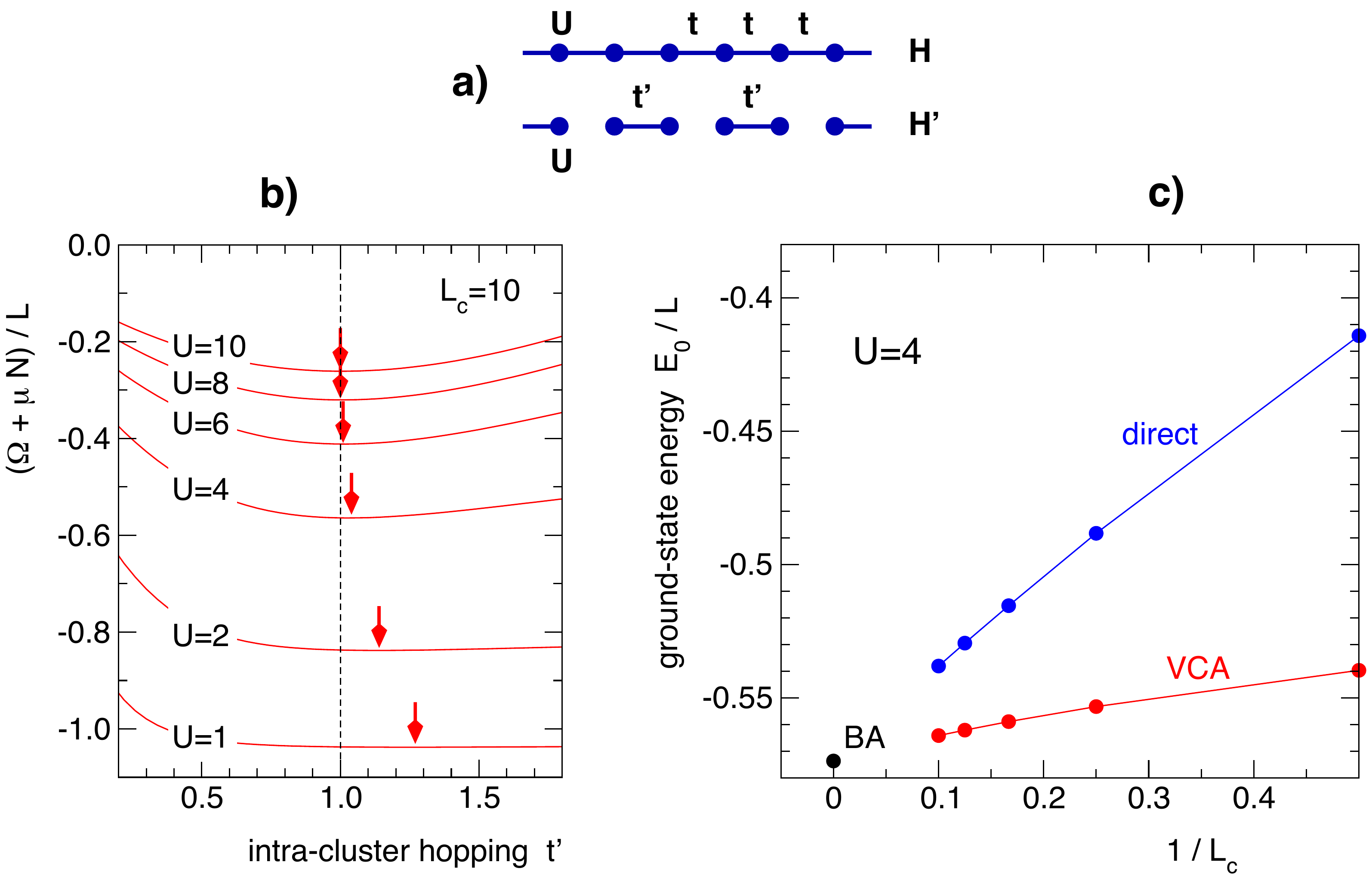}
\caption{
(adapted from Ref.\ \cite{BHP08}).
a) Original system: one-dimensional Hubbard model. Reference system: decoupled clusters (for $L_{\rm c}=2$). 
b) SFT grand potential per site and shifted by $\mu N$ as a function of the intra-cluster nearest-neighbor hopping $t'$. 
VCA calculation (with $L_{\rm c}=10$) for the one-dimensional Hubbard model at zero temperature, half-filling and different $U$ as indicated. 
The nearest-neighbor hopping $t=1$ sets the energy scale.
Arrows mark the stationary points.
c) VCA ground-state energy per site as a function of $1/L_{\rm c}$ for $U=4$ at the respective stationary points compared with the corresponding results for an isolated cluster and the exact results known from the Bethe ansatz (BA) \cite{LW68}.
}
\label{fig:hopp}
\end{figure}

In the first example \cite{BHP08} we will consider the one-dimensional Hubbard model at zero temperature and half-filling with hopping $t=1$ between nearest neighbors, see Fig.\ \ref{fig:hopp}a.
A tiling of the one-dimensional lattice into ``clusters'' is particularly simple: Each cluster is a finite chain of $L_{\rm c}$ sites. 
We treat the intra-cluster nearest-neighbor hopping $t'$ as the only variational parameter. 
This is the most obvious choice. Nevertheless, one may numerically check that the optimal on-site hopping $t'_{ii, \rm opt}=t_{ii}=0$.
The same holds for the hopping between second nearest neighbors: $t'_{2\rm-nd, \rm opt}=0$.
Again this is predicted by particle-hole symmetry. 
On the other hand, if a third nearest-neighbor hopping is introduced as a variational parameter, it acquires a small finite value at the stationary point.
Interestingly, one also finds $t'_{\rm pbc, opt}=0$ \cite{PAD03} where $t'_{\rm pbc}$ is a hopping parameter that links the two edge sites of the cluster with each other. $t'_{\rm pbc} = t'$ would be a realization of periodic boundary conditions but the calculation shows that open boundaries, $t'_{\rm pbc, opt}=0$, are preferred.
Furthermore, one may also relax the constraint that the hopping $t'$ be the same for all pairs of nearest neighbors.
In this case one finds the strongest deviations close to the edges of the reference systems \cite{BHP08}.

Fig.\ \ref{fig:hopp}b shows the dependence of the SFT grand potential $\Omega[\ff \Sigma_{t'}]$ on the {\em single} variational parameter $t'$. 
Actually, $(\Omega + \mu \langle N \rangle)/L$ is plotted. 
At zero temperature and at the stationary point, this is the (approximate) ground-state energy of the Hubbard model per site.
There is a stationary point, a minimum in this case, with the optimal value for the intra-cluster hopping $t'_{\rm opt}$ being close to but different from the physical value $t=1$ for strong Coulomb interaction $U$. 
Note that the CPT is given by $t' = t$ and that there is a gain in binding energy due to the optimization of $t'$, namely $\Omega(t'_{\rm opt}) < \Omega(t)$ which implies that the VCA improves on the CPT result.

It is physically reasonable that in case of a stronger interaction and thus more localized electrons, switching off the inter-cluster hopping is less significant and must therefore be outweighed to a lesser degree by an increase of the intra-cluster hopping. 
A considerably large deviation from the physical hopping, $t'_{\rm opt}>t$, is only found for the weakly interacting system. 
However, even a ``strong'' approximation of the self-energy (measured as a strong deviation of $t'_{\rm opt}$ from $t$) becomes irrelevant in the weak-coupling limit as the self-energy becomes small.
With decreasing $U$, the self-energy functional becomes flatter and flatter until at $U=0$ the $t'$ dependence is completely irrelevant.
Note that not only the non-interacting limit but also the atomic limit ($t=0$) is exactly reproduced by the VCA.
In latter case, the reference system becomes identical with the original system at $t'=0$.

Fig.\ \ref{fig:hopp}c shows the VCA ground-state energy (per site) at a fixed interaction strength $U=4$ as a function of the inverse cluster size $1/L_{\rm c}$.
By extrapolation to $1/L_{\rm c} = 0$ one recovers the exact Bethe-Ansatz result (BA) \cite{LW68}. 
Furthermore, the VCA is seen to improve the ground-state energy as compared to calculations done for an isolated Hubbard chain with open boundaries. 
Convergence to the BA result is clearly faster within the VCA. 
Note that, opposed to the VCA, the direct cluster approach is not exact for $U=0$.

\begin{figure}[t]
\centering
\includegraphics[width=0.5\columnwidth]{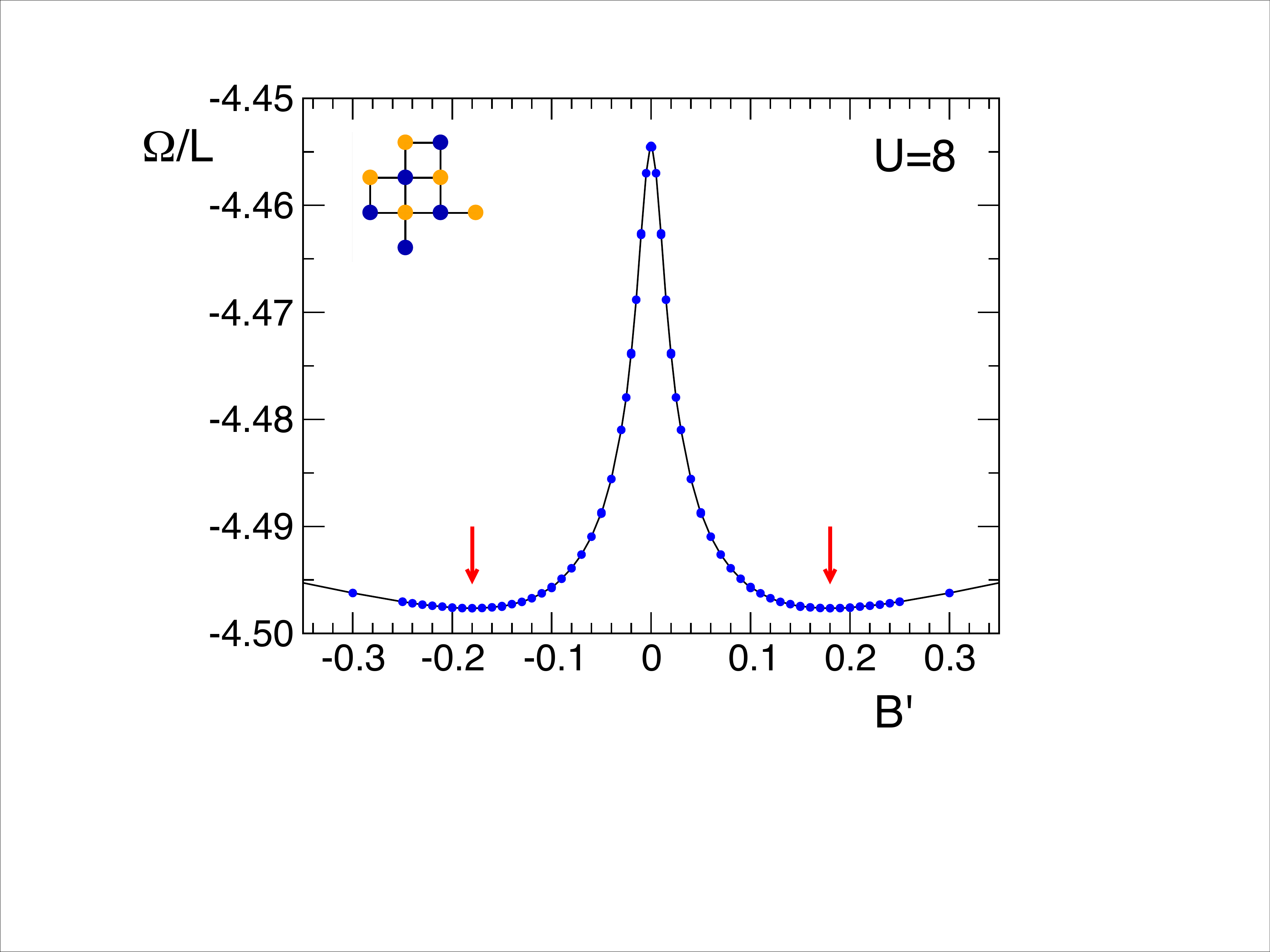}
\caption{
(adapted from Ref.\ \cite{DAH+04}).
SFT grand potential per site as a function of the strength of a fictitious staggered magnetic field $B'$.
Calculation for the two-dimensional half-filled Hubbard model on the square lattice at zero temperature and $U=8$, $t=1$.
The reference system consists of disconnected clusters with $L_{\rm c}=10$ sites each, see the inset for the cluster geometry.
Arrows mark the two equivalent stationary points. 
}
\label{fig:afm}
\end{figure}

\subsection*{Antiferromagnetism}

\index{antiferromagnetism}

With the second example \cite{DAH+04} we return to our original motivation, see Fig.\ \ref{fig:field}. 
One of the main drawbacks of the CPT consists in its inability to describe spontaneous symmetry breaking. 
Consider SU(2) transformations in spin space and antiferromagnetic order, for example. 
As the exact solution of a finite Hubbard cluster is necessarily spin symmetric, i.e.\ invariant under SU(2) transformations, and as the CPT equation proliferates this symmetry, the antiferromagnetic order parameter, the staggered magnetization $m$, must always be zero if there is no physically applied staggered magnetic field which would explicitly break the symmetry, i.e.\ if $B=0$. 
The VCA, on the other, in principle allows for a spontaneous SU(2) symmetry breaking. 
Namely, treating an intra-cluster fictitious staggered magnetic field of strength $B'$ as a variational parameter offers the possibility for a symmetry-broken stationary point with $B'_{\rm opt} \ne 0$.
The reference-system Hamiltonian is given by
\be
  H' = H'\big|_{B'=0} - B' \sum_{i\sigma} z_i (n_{i\uparrow} - n_{i\downarrow}) \; ,
\ee
where $z_i = +1$ for sites on sublattice A, and $z_i = -1$ for sublattice B (yellow and blue sites in the inset of Fig.\ \ref{fig:afm})

The main part of Fig.\ \ref{fig:afm} displays VCA results for the half-filled two-dimensional Hubbard model on the square lattice at zero temperature. 
Decoupled clusters with $L_{\rm c}=10$ sites are considered as a reference system, and the staggered field $B'$ is the only variational parameter considered.
There is a stationary point at $B'=0$ which corresponds to the paramagnetic phase and to the CPT.
In addition, however, there are two equivalent stationary points at {\em finite} $B'$ corresponding to a phase with antiferromagnetic order. 
Comparing the ground-state energies of both the antiferromagnetic and the paramagnetic phase, shows that the former is thermodynamically stable.

One should be aware, however, that the VCA, as any cluster mean-field approach, tends to overestimate the tendency towards magnetic order. 
Furthermore, a finite-temperature calculation is expected to produce a finite order parameter for the two-dimensional but also for the one-dimensional case which would be at variance with the Mermin-Wagner theorem \cite{MW66}.
What is missing physically in the VCA is the effect of long wave-length spin excitations. 
The VCA is therefore restricted to cases where the physical properties are dominated by short-range correlations on a scale accessible by an exactly solvable finite cluster.

\subsection*{Mott metal-insulator transition}

\index{Mott transition}

The third example \cite{BKS+09} addresses a first-order (discontinuous) phase transition.
This type of phase transitions can be studied conveniently within the SFT framework as there is an explicit expression for a thermodynamical potential available from the very beginning.
We again consider the two-dimensional Hubbard model on the square lattice at half-filling and zero temperature but disregard the antiferromagnetic phase and enforce a paramagnetic state by choosing $B'=0$. 
The paramagnetic system is expected to undergo a transition from a correlated metal at weak $U$ to a Mott insulator at strong $U$. 
This Mott transition is first of all interesting from a fundamental point of view as it is driven by electronic correlations opposed to other types of metal-insulator transitions \cite{Geb97}.
The Mott insulator is characterized by a gap of the order of $U$ in the single-particle excitation spectrum which is only weakly dependent on temperature.
One therefore expects that a possible metal-insulator quantum phase transition at zero temperature is of relevance for the high-temperature state of the system as well where it should give rise to a smooth crossover between a more metallic and a more insulating state.
The crossover takes place at temperatures which may be well above the N\'eel temperature where the system is paramagnetic.
This is the motivation to ignore the magnetic phase for the zero-temperature calculation.

\begin{figure}[t]
\centering
\includegraphics[width=0.6\columnwidth]{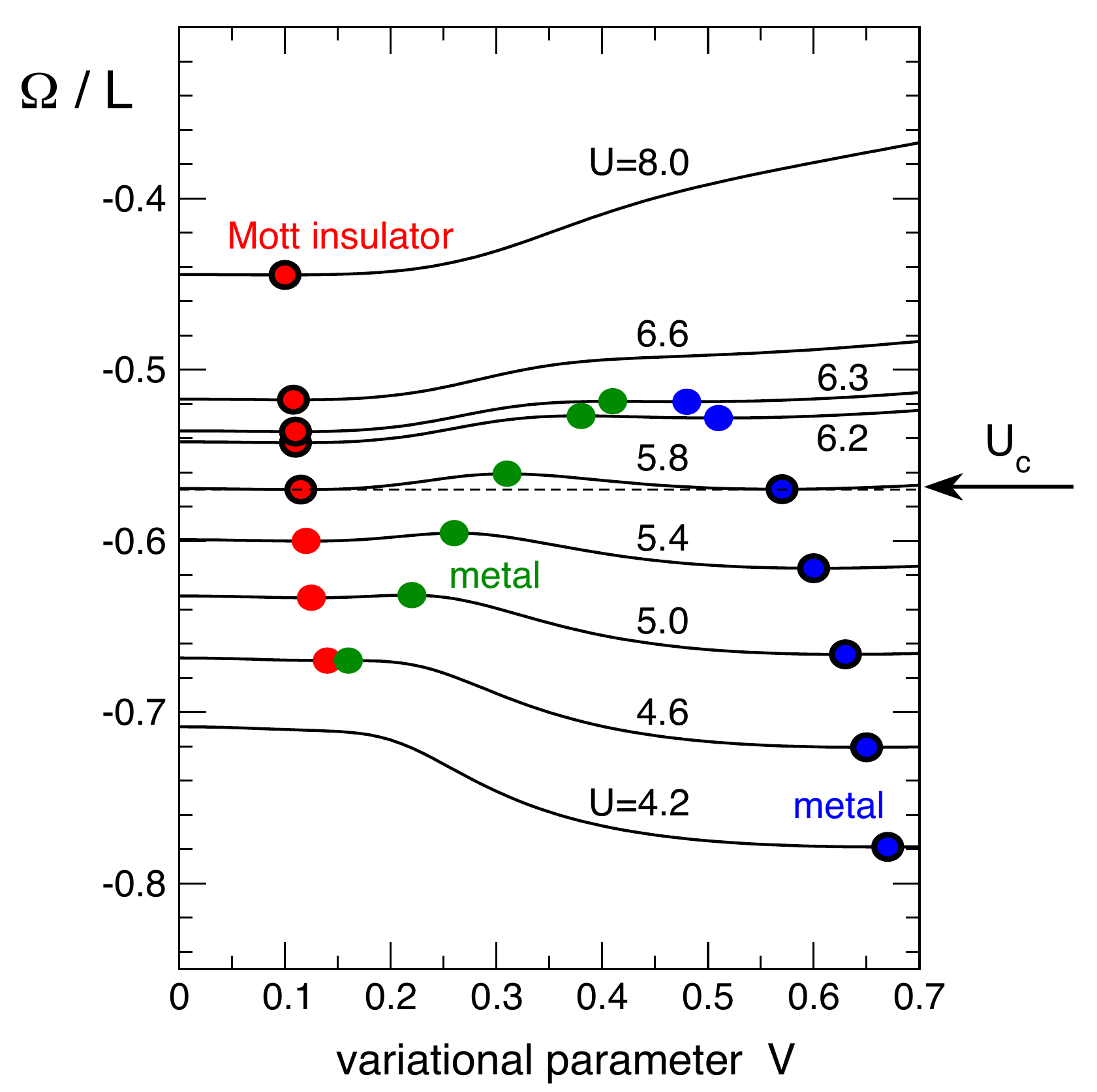}
\hspace{0.1\textwidth}
\includegraphics[width=0.2\columnwidth]{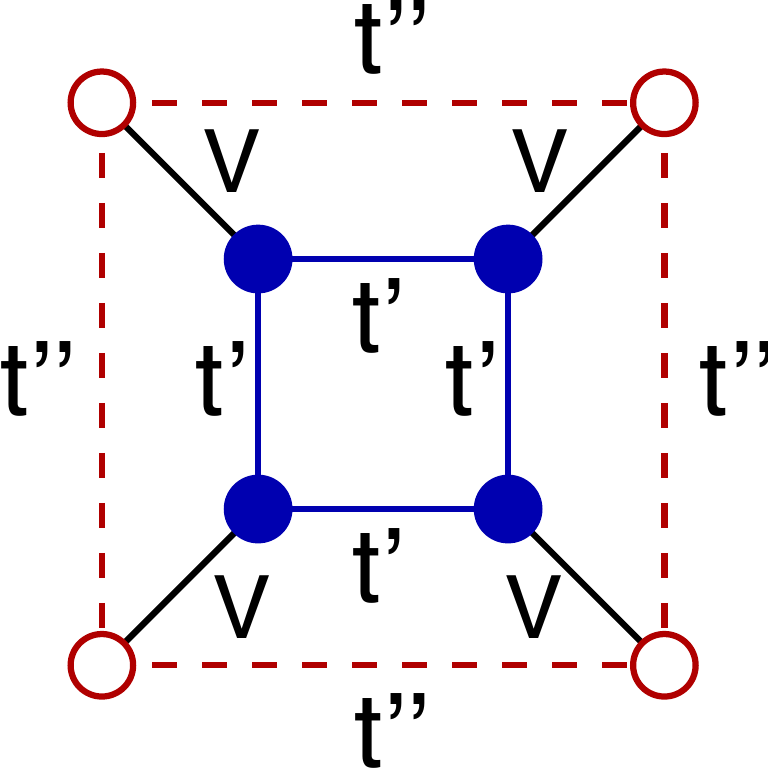}
\caption{
(adapted from Ref.\ \cite{BKS+09}).
{\em Left:} SFT grand potential per site as a function of the variational parameter $V$. 
VCA calculation for the two-dimensional Hubbard model on the square lattice at zero temperature, half-filling and different $U$ as indicated.
The nearest-neighbor hopping $t=1$ sets the energy scale.
Symbols: stationary points. 
Red: Mott insulator, green and blue: metal.
Fat symbols: thermodynamically stable phase. 
The first-order Mott transition is marked by an arrow.
{\em Right:} Sketch of the building block of the reference system. 
Blue filled dots: correlated sites with $U>0$. 
Red open dots: bath sites with $U=0$.
Calculations for $t'=t$, $t''=0$, arbitrary $V$.
}
\label{fig:mott}
\end{figure}

Fig.\ \ref{fig:mott} (right) shows the building block of the reference system. 
This is a cluster with $L_{\rm c}=4$ correlated sites (filled blue dots) but with four bath sites (open red dots) in addition, i.e.\ there are $n_{\rm s}=2$ {\em local} degrees of freedom (one additional bath site per correlated site).
As mentioned above, including bath sites in a VCA calculation improves upon the description of local correlations.
This is an important ingredient to understand the Mott transition: 
A paradigmatic picture of the Mott transition could be worked out in the limit of infinite spatial dimensions with the help of the dynamical mean-field theory \cite{MV89,GKKR96}. 
In this limit the self-energy becomes a completely local quantity, $\Sigma_{ij\sigma}(i\omega_{n}) = \delta_{ij} \Sigma_{i\sigma}(i\omega_{n})$ \cite{MH89b}, and therefore the local, temporal degrees of freedom ($n_{\rm s}$) are dominating over the spatial ones ($L_{\rm c}$). 
For two dimensions, the considered reference system with $L_{\rm c}=4$ and $n_{\rm s}=2$ is expected to represent a good compromise between importance of local and non-local correlations and to result in a reasonable approximation. 

VCA calculations with the full set of variational parameters indicated in Fig.\ \ref{fig:mott} (right) have shown \cite{BKS+09} that the hopping between the correlated and the bath sites $V$ is the most important parameter to be optimized while $t'_{\rm opt}\approx t$ and $t''_{\rm opt}\approx 0$ can safely be ignored, i.e.\ set to the {\em a priori} plausible values $t'_{\rm opt} = t$ and $t''_{\rm opt} = 0$.
The on-site energies of the correlated and of the bath sites are fixed by particle-hole symmetry anyway.
This drastically simplifies the study as the SFT grand potential $\Omega[\ff \Sigma_{V}]$ can be regarded as a function of a single variational parameter $V$ only. 

Fig.\ \ref{fig:mott} (left) displays the SFT grand potential, shifted by $\mu N$, per correlated site as a function of $V$ for different $U$. 
For weak interactions $U<U_{c2}\approx 6.35$ there is a stationary point (a minimum) at a comparatively large $V_{\rm opt}$ which describes a metallic phase (blue dots). 
The metallic character of the phase can be inferred from the finite value of the imaginary part of the local Green's function $\mbox{Im}\,G_{ii\sigma}(i\omega)$ for $\omega \to 0$ (see Ref.\ \cite{BKS+09}). 
Above the critical value $U_{c2}$ no metallic solution can be found. 
There is, however, an insulating phase for strong $U$ (red dots). 
The respective stationary point (a minimum) of the SFT grand potential is found at a comparatively low value for $V_{\rm opt}$ and can be traced with decreasing $U$ down to another critical value $U_{c1} \approx 4.6$.
For $U<U_{c1}$ there is no insulating phase. 

It is interesting to observe that in the regime $U_{c1}<U<U_{c2}$ the metallic and the insulating phase are coexisting, i.e.\ that there are two stationary points of the grand potential. 
There is actually a third stationary point in the coexistence region, indicated by the green dots, where the SFT grand potential is at a maximum. 
Note that any stationary point, minimum, maximum or saddle point (in higher-dimensional parameter spaces), must be considered as an admissible solution within the SFT. 
However, the grand potential of the third phase is always higher than the grand potentials of the other phases.
It therefore describes a physically irrelevant metastable phase but mathematically explains why the other two phases cease to exist above (below) a certain critical interaction $U_{c2}$ ($U_{c1}$). 

For a given $U$, the phase with the lowest grand potential is thermodynamically stable (see fat symbols). 
This means that the system is a correlated metal for $U<U_{c}$ and a Mott insulator for $U>U_{c}$ where the critical value for the Mott transition $U_{c} \approx 5.8$ is given by the interaction strength for which the metal and insulator have the same grand potential (the same ground-state energy at zero temperature), see the arrow in Fig.\ \ref{fig:mott}.
Therefore, at $U_{c}$ the optimal hopping parameter $V_{\rm opt}$ {\rm jumps} between the large metallic and the small insulating value. 
Consequently, the ground state and thus the self-energy of the reference system changes abruptly at $U_{c}$. 
This leads to a discontinuous change of the SFT Green's function as well as of all observables that are computed as derivatives of the (optimized) SFT grand potential. 
The phase transition is of {\em first order} or {\em discontinuous}. 
This is interesting since the Mott transition in the Hubbard model on an infinite-dimensional, e.g.\ hyper-cubic, lattice is known \cite{GKKR96} to be of second order or continuous.
The VCA calculation discussed here actually corrects a mean-field artifact that is due to the neglect of non-local short-range antiferromagnetic correlations (see Ref.\ \cite{BKS+09} for an extended discussion).
For a discussion of more recent developments see Ref.\ \cite{SGR+14}, for example.

\section{Relation to other methods and conclusions}

\index{cluster mean-field theories}
\index{dynamical impurity approximation}
\index{cellular dynamical mean-field theory}

Concluding, it is an appealing idea to divide a correlated lattice-fermion problem into small isolated clusters for which the problem can be solved easily and in a second step to employ the decoupled-cluster solution to construct the solution for the original lattice model.
This construction must be approximate and has the spirit of a cluster mean-field theory, where the intra-cluster correlations are treated in a much better and explicit way than the inter-cluster correlations. 
We have learnt that this construction cannot be based on a technique that uses many-body wave-functions, one rather has to employ Green's functions, namely the single-particle Green's function or, equivalently, the self-energy, as it is done with the cluster-perturbation theory. 

The self-energy-functional theory conceptually improves the CPT in several respects: 
First, it removes the arbitrariness that is inherent to the CPT regarding the choice of the cluster parameters.
Second, it introduces an element of self-consistency or variational character by which it becomes possible to study phases which possess a symmetry different from the symmetry of the isolated cluster, i.e.\ one can address spontaneously symmetry-broken phases (collective magnetism, superconductivity etc.).
Third, the SFT provides us with an explicit, though approximate, expression for a thermodynamical potential from which all observables have to be derived. This ensures that the approach is consistent in itself and obeys general thermodynamical relations, an important point that is missing in the plain CPT as well. 

\begin{figure}[t]
\centering
\includegraphics[width=0.6\columnwidth]{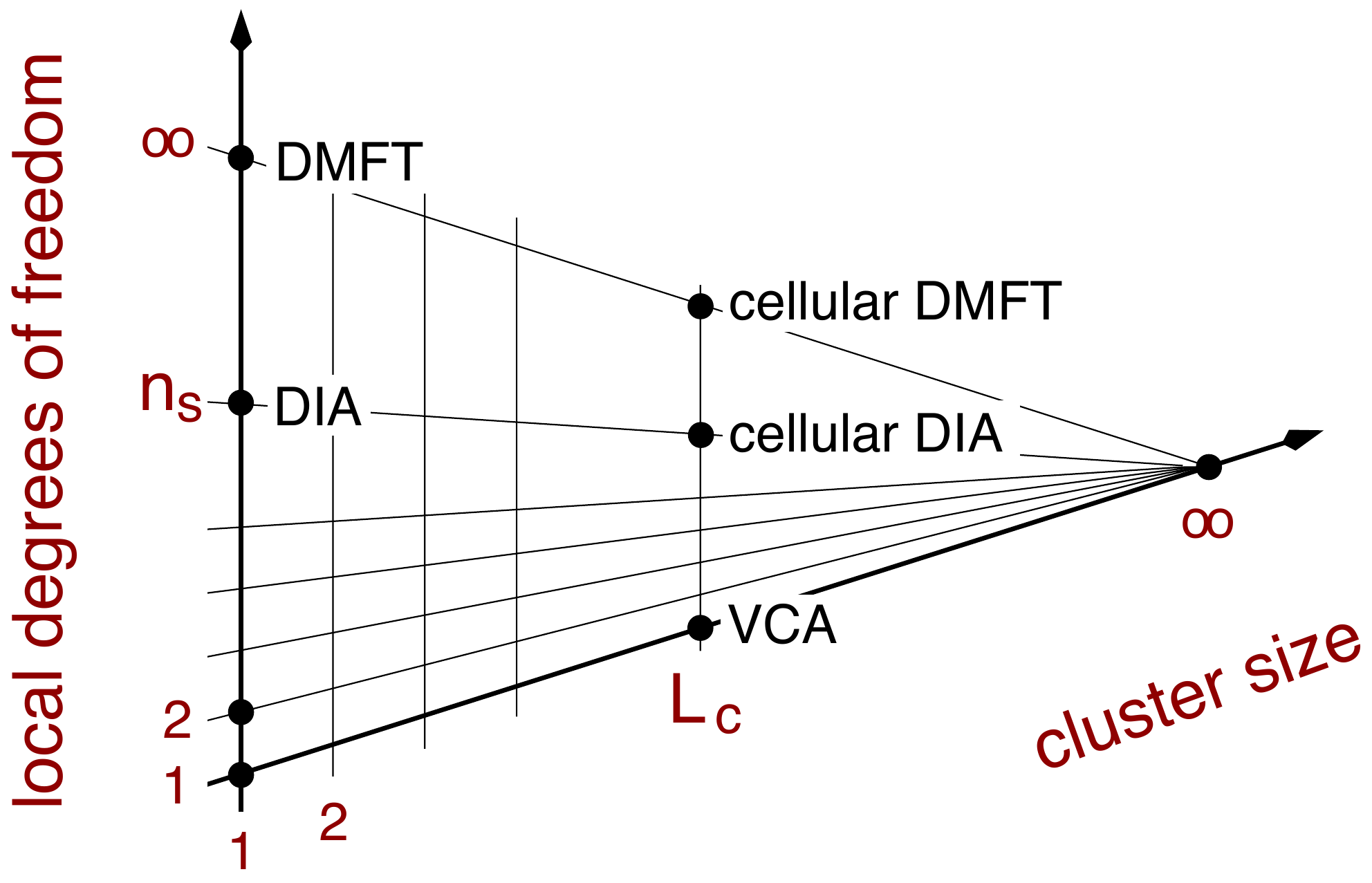}
\caption{
Schematic picture for the relation between different approximations that can be constructed within the self-energy-functional theory. 
See text for discussion.
}
\label{fig:ap}
\end{figure}

The self-energy-functional theory should actually be seen as a theoretical frame that allows to construct different approximations.
Each approximation is characterized by the choice of a corresponding reference system. 
Typically, this consists of decoupled clusters with a number of $L_{\rm c}$ correlated sites each and an additional number of $n_{\rm s}-1$ uncorrelated bath sites. 
A large number $L_{\rm c}$, i.e.\ large clusters are necessary to include short-range correlations as good as possible, and a large number of local degrees of freedom $n_{\rm s}$ is recommendable to improve the description of local, temporal correlations.
There is the ubiquitous tradeoff between the quality of the approximation on the one hand and the numerical effort on the other as the problem must be exactly solved for the isolated cluster. 
Using an exact-diagonalization solver the effort roughly increases exponentially with $L_{\rm c}$ and $n_{\rm s}$.
The ``space'' of possible approximations that is spanned by $L_{\rm c}$ and $n_{\rm s}$ is sketched with Fig.\ \ref{fig:ap}.

The most simple approximation is given by $L_{\rm c}=1$ and $n_{\rm s}=1$. 
Here, one approximates the self-energy of the lattice model by the self-energy of the atomic problem. 
This is in the spirit of the Hubbard-I approximation \cite{Hub63}.
For $L_{\rm c}>1$ we find the variational cluster approximation that we have discussed at length here.
Obviously, one would recover the exact solution of the lattice model in the limit $L_{c}\to\infty$.

Choosing a ``cluster'' with a single correlated site only, $L_{\rm c}=1$, but introducing a number of bath sites $n_{\rm s}-1\ge 1$ specifies another approximation which is called dynamical impurity approximation (DIA). 
This is a true mean-field approximation as all non-local two-particle spatial correlations are neglected in this case.

\index{dynamical mean-field theory}

Obviously, an ideal embedding of a single site into the lattice or an ideal mean-field theory is realized with an infinite number of bath sites $n_{\rm s}\to \infty$.
In this case, all local, temporal correlations are treated exactly -- opposed to a {\em static} mean-field theory like the Hartree-Fock approach.
This optimal mean-field theory turns out to be identical with the well-known dynamical mean-field theory (DMFT) \cite{MV89,GKKR96}.
In fact, one immediately recognizes that the reference system in this case is just the single-impurity Anderson model (if one starts out from a single-band Hubbard model, for example).
Within DMFT the parameters of this impurity model are fixed by imposing the so-called DMFT self-consistency condition, stating that the impurity Green's function of the impurity model should be equal to the local Green's function of the lattice model. 
This is realized by setting up a self-consistency scheme which requires an iterated solution of the impurity model. 
Within the SFT, on the other hand, the parameters of the reference impurity system are fixed by the SFT Euler equation which is the condition for the stationarity of the SFT grand potential.
In fact both, the DMFT self-consistency equation and the SFT Euler equation, are basically equivalent in this case. 
From \refeq{sfp}, we have:
\be
\frac{\partial}{\partial \ff t'} \Omega[\ff \Sigma_{\ff t'}]
=
\frac{1}{\beta}
\sum_{n,ij\sigma}
\left(
\frac{1}{\ff G_{0}^{-1}(i\omega_{n}) - \ff \Sigma(i\omega_{n})} - \ff G'(i\omega_{n})
\right)_{ij\sigma}
\frac{\partial \Sigma_{ji\sigma}(i\omega_{n})}{\partial \ff t'}
= 0 \; ,
\labeq{dmft}
\ee
where $\ff G'$ denotes the Green's function of the reference system.
As the self-energy of a single-impurity Anderson model is local, $\Sigma_{ij\sigma}(i\omega_{n})  = \delta_{ij} \Sigma_{i\sigma}(i\omega_{n})$, and non-zero on the correlated impurity site only, this SFT Euler equation is satisfied if the {\em impurity} Green's function $G'_{\rm imp}(i\omega_{n})$ is equal to the (approximate) {\em local} Green's function of the lattice model, i.e.\ if the local elements of the bracket vanish. 
This is the DMFT self-consistency equation. 

\index{dynamical impurity approximation}

What happens if $n_{\rm s}<\infty$?
For a small number of bath sites an exact solution of the reference system by means of exact-diagonalization techniques becomes feasible in practice. 
The resulting approximation, the DIA, differs from DMFT as the bracket in \refeq{dmft} will never vanish in this case. 
This is easily seen by noting that the Green's function of an $n_{\rm s}<\infty$ impurity model has a finite number of poles on the real frequency axis while the approximate lattice Green's function inherits its analytical structure from the non-interacting lattice Green's function $\ff G_{0}$ which, for an infinite lattice, may exhibit isolated poles but must have branch cuts as well. 
The DIA is different and actually inferior compared with the DMFT but does not need an advanced ``solver'' (note that solving the single-impurity Anderson model with $n_{\rm s}=\infty$ is still a demanding many-body problem).
It has turned out, however, that with a few bath sites only, one often has a rather reliable approach to study the thermodynamics of a lattice model (see Ref.\ \cite{Pot03b}, for an example).
This is also known from the exact-diagonalization approach to DMFT \cite{CK94}.
DMFT-ED considers the impurity model with a small $n_{\rm s}$, as is done in the DIA, but employs another, actually more {\em ad hoc} condition to fix the parameters of the impurity model, namely one minimizes a suitably defined ``distance'' between the two local elements of the Green's function in the DMFT self-consistency condition to satisfy this at least approximately. 
The approach is able to yield similar results as the DIA in practice and more easily implemented numerically but lacks internal consistency. 

\index{cluster mean-field theories}

The third example discussed in the previous section has shown that one can favorably make use of approximations where a small number of bath degrees of freedom $n_{\rm s}>1$ are combined with a cluster approach $L_{\rm c}>1$. 
This approach is a VCA with additional bath sites and may be termed ``cellular DIA'' (see Fig.\ \ref{fig:ap}) since it is related to the cellular DMFT \cite{KSPB01,LK00} in the same way as the DIA is related to the DMFT. 
Note that with increasing cluster size $L_{\rm c}\to \infty$, all approaches, the VCA, the cellular DIA as well as the cellular DMFT must recover the exact solution of the lattice problem in principle.

\index{cellular dynamical mean-field theory}

\index{dynamical cluster approximation}

Another prominent and widely used cluster mean-field theory is the dynamical cluster approximation (DCA) \cite{HTZ+98}.
As compared to the cellular DMFT, this is a cluster extension of dynamical mean-field theory that avoids one of the main drawbacks of various cluster approaches, namely the artificial breaking of the translational symmetries. 
Already the CPT yields an approximate Green's function that merely reflects the translational symmetries of the ``superlattice'', which periodically repeats the basic cluster, rather than the symmetries of the underlying physical lattice.
Contrary, the DCA provides a Green's function and a self-energy with the correct symmetries but, on the other hand, must tolerate that the self-energy is discontinuous as a function of $\ff k$ in the reciprocal space.
Here, we briefly mention that it is possible to re-derive the DCA within in the framework of the SFT as well.
This is carried out in detail in Ref.\ \cite{PB07} and is based on the idea that, with a proper modification of the hopping parameters the {\em original} lattice model, $\ff t \mapsto \widetilde{\ff t}$, which becomes irrelevant for $L_{\rm c}\to \infty$, the DCA becomes equivalent with the cellular DMFT.
In a similar way \cite{KD05} another cluster-mean-field variant, the periodized cellular DMFT \cite{BPK04} can be re-derived within the SFT.

To summarize, the self-energy-functional approach not only recovers a number of well-known mean-field and cluster mean-field concepts and provides a unified theoretical framework to classify the different approaches but has also initiated the construction of new non-perturbative and consistent approximations, the most prominent example of which is the variational cluster approximation. 
The challenges for future developments are manifold: 
Let us mention only two directions here: 
The first consists in the generalization of the SFT to many-body lattice models far away from thermal equilibrium \cite{HEAP13}. 
This requires a reformulation of the theory in terms of non-equilibrium Green's functions but offers the exciting perspective to study the real-time dynamics of strongly correlated systems in a non-perturbative and consistent way.
Another equally important direction of future work consists in an extension of the theory to correlated lattice models with non-local and even long-ranged interactions. 
First promising steps have already been made \cite{Ton05}.
The restriction to local Hubbard-type interactions, inherent to all of the approaches mentioned here (see Fig.\ \ref{fig:ap}), represents an  eventually unacceptable model assumption which must be abandoned.

\subsection*{Acknowledgement}

Support of this work by the Deutsche Forschungsgemeinschaft through FOR 1346 is gratefully acknowledged.


\clearchapter

\end{document}